\documentclass[traditabstract]{aa} 
\usepackage{txfonts}
\usepackage{graphicx}
\usepackage{longtable}

\begin{document}

\title{Orbital solutions of eight close sdB binaries and constraints on the nature of the unseen companions}

\author{S.~Geier \inst{1}
   \and R.~H.~\O stensen \inst{2}
   \and U.~Heber \inst{3}
   \and T.~Kupfer \inst{4}
   \and P.~F.~L.~Maxted \inst{5}
   \and B.~N.~Barlow \inst{6}
   \and M.~Vu\v ckovi\'c \inst{7}
   \and A.~Tillich \inst{3}
   \and S.~M\"uller \inst{3}
   \and H.~Edelmann \inst{3}
   \and L.~Classen \inst{3}
   \and A.~F.~McLeod \inst{2,4}}

\offprints{S.\,Geier,\\ \email{geier@sternwarte.uni-erlangen.de}}

\institute{European Southern Observatory, Karl-Schwarzschild-Str.~2, 85748 Garching, Germany
\and Institute of Astronomy, KU Leuven, Celestijnenlaan 200D, B-3001 Heverlee, Belgium
\and Dr. Karl Remeis-Observatory \& ECAP, Astronomical Institute,
Friedrich-Alexander University Erlangen-Nuremberg, Sternwartstr. 7, D 96049 Bamberg, Germany
\and Department of Astrophysics/IMAPP, Radboud University Nijmegen, P.O. Box 9010, 6500 GL Nijmegen, The Netherlands
\and Astrophysics Group, Keele University, Staffordshire, ST5 5BG, UK
\and Department of Physics, High Point University, 833 Montlieu Avenue, High Point, NC 27262, USA
\and European Southern Observatory, Alonso de Cordova 3107, Vitacura, Casilla 19001, Santiago, Chile
}

\date{Received \ Accepted}

\abstract{The project Massive Unseen Companions to Hot Faint Underluminous Stars from SDSS (MUCHFUSS) aims at finding hot subdwarf stars (sdBs) with massive compact companions such as white dwarfs, neutron stars, or stellar-mass black holes. In a supplementary programme we obtained time-resolved spectroscopy of known hot subdwarf binary candidates. Here we present orbital solutions of eight close sdB binaries with orbital periods ranging from $\sim0.1\,{\rm d}$ to $10\,{\rm d}$, which allow us to derive lower limits on the masses of their companions. Additionally, a dedicated photometric follow-up campaign was conducted to obtain light curves of the reflection-effect binary HS\,2043+0615. We are able to constrain the most likely nature of the companions in all cases but one, making use of information derived from photometry and spectroscopy. Four sdBs have white dwarf companions, while another three are orbited by low-mass main sequence stars of spectral type M.\\ 

}

\maketitle

\section{Introduction \label{s:intro}}

Subluminous B stars or hot subdwarfs (sdBs) are core helium-burning stars with thin hydrogen envelopes and masses around $0.5\,M_{\rm \odot}$ 
(Heber \cite{heber86}, see Heber \cite{heber09} for a review). A large proportion of the sdB stars ($40\,\%$ to $80\,\%$) are members of short-period binaries (Maxted et al. \cite{maxted01};  Napiwotzki et al. \cite{napiwotzki04a}). Several studies aimed at determining the orbital parameters of short-period sub\-dwarf binaries and have found periods ranging from $0.05\,{\rm d}$ to more than $10\,{\rm d}$ with a peak around $0.5$ to $1.0\,{\rm d}$ (e.g. Morales-Rueda et al. \cite{morales03}; Edelmann et al. \cite{edelmann05}; Copperwheat et al. \cite{copperwheat11}). For these close binary sdBs, common envelope (CE) ejection is the only feasible formation channel. At first, two main sequence stars evolve in a binary system. The more massive one will then enter the red-giant phase and eventually fill its Roche lobe. Triggered by dynamically unstable mass transfer, a common envelope is formed. Owing to friction the two stellar cores lose orbital energy, which is deposited within the envelope and leads to a shrinking of the binary orbit. Eventually, the common envelope is ejected and a close binary system is formed, which contains a core helium-burning sdB and a main sequence companion. A close sdB binary with white dwarf (WD) companion is formed after two consecutive phases of mass-transfer (Han et al. \cite{han02,han03}). 

The nature of the close companions to sdB stars is hard to constrain in general, since most of those binaries are single-lined with the hot subdwarf being the only star detectable in the spectrum. Measuring the Doppler reflex motion of this star from time-resolved spectra, the radial velocity (RV) curve can be derived and a lower limit can be given for the mass of the companion from the binary mass function. These lower limits are in general compatible with main sequence stars of spectral type M or compact objects such as white dwarfs.  

Subdwarf binaries with massive WD companions are candidates for supernova type Ia (SN Ia) progenitors because these systems lose angular momentum through the emission of gravitational waves and start mass transfer. This mass transfer, either from accretion of helium onto the WD during the sdB phase (see Wang et al. \cite{wang13} and references therein), or the subsequent merger of the system (Tutukov \& Yungelson \cite{tutukov81}; Webbink \cite{webbink84}), may cause the companion to explode as SN Ia. Two of the best known candidate systems for SN\,Ia are sdB+WD binaries (Maxted et al. \cite{maxted00}; Geier et al. \cite{geier07}; Vennes et al. \cite{vennes12}; Geier et al. \cite{geier13}). 
More candidates, some of which might even have more massive compact companions (i.e. neutron stars or black holes), have been found as well (Geier et al. \cite{geier08}, \cite{geier10a}, \cite{geier10b}). Such systems are also predicted by binary evolution theory (Podsiadlowski et al. \cite{podsi02}; Pfahl et al. \cite{pfahl03}; Yungelson \& Tutukov \cite{yungelson05};  Nelemans \cite{nelemans10}).

The project Massive Unseen Companions to Hot Faint Underluminous Stars from SDSS (MUCHFUSS) aims at finding sdBs with such massive compact companions. We selected and classified hot subdwarf stars from the Sloan Digital Sky Survey (SDSS, Data Release 7, Abazajian et al. \cite{abazajian09}) by colour selection and visual inspection of their spectra. Radial velocity variable subdwarfs with high shifts were selected as candidates for time-resolved spectroscopy to derive their orbital para\-meters and follow-up photometry to search for features such as eclipses in their light curves. 

Target selection and follow-up strategy were presented in Geier et al. (\cite{geier11a,geier12}). In a spin-off project the kinematics of fast-moving sdBs in the halo have been studied (Tillich et al. \cite{tillich11}). The spectroscopic and photometric follow-up campaigns of the binary candidates are described in Geier et al. (\cite{geier11b}), Kupfer et al. (\cite{kupfer13}), and Schaffenroth et al. (\cite{schaffenroth13b}). We discovered three eclipsing binary systems, two of them with brown dwarf companions (Geier at al. \cite{geier11c}; Schaffenroth et al. \cite{schaffenroth13a}), and one hybrid sdB pulsator with reflection effect (\O stensen et al. \cite{oestensen13}). Here we report on our supplementary programme that investigates known hot subdwarf binaries.

\section{MUCHFUSS supplementary programme \label{s:backup}}

In addition to the priority objects, the MUCHFUSS project targeted known sdB binaries of special importance, whenever scheduling constraints or weather conditions were unsuitable to execute the main programme. In particular, objects were included for which light variations either caused by eclipses, by the reflection effect, or by stellar oscillations have been reported in the literature, because this complementary information is of great value for understanding their nature and evolutionary history. 

Providing sufficient RV information is therefore rewarding. We also keep a list of targets, that have insufficient RV coverage, mostly from the SPY survey (Lisker et al. \cite{lisker05}) and Copperwheat et al. (\cite{copperwheat11}). The highlight of our supplementary programme so far was the discovery of the ultracompact sdB+WD binary CD$-$30$^\circ$11223. It is not only the shortest-period ($P\simeq0.049\,{\rm d}$) hot subdwarf binary known, but also an excellent progenitor candidate for an underluminous SN\,Ia (Geier et al. \cite{geier13}). 

Here we present orbital solutions of eight close hot subdwarf binaries, which allow us to derive lower limits on the masses of their companions. Furthermore, we are able to constrain the most likely nature of the companions in all cases but one, making use of additional information derived from photometry and spectroscopy. 

\section{Observations and data reduction}

\subsection{Spectroscopic observations}\label{s:data}
 
Follow-up medium resolution spectra were taken during de\-dicated MUCHFUSS follow-up runs (Geier et al. \cite{geier11a,geier11b}; Kupfer et al. \cite{kupfer13}) with the EFOSC2 spectrograph ($R\simeq2200,\lambda=4450-5110\,{\rm \AA}$) mounted at the ESO-NTT, the ISIS spectrograph ($R\simeq4000,\lambda=3440-5270\,{\rm \AA}$) mounted at the WHT, the TWIN spectrograph mounted at the CAHA-3.5m telescope ($R\simeq4000, \lambda=3460-5630\,{\rm \AA}$), and the Goodman spectrograph mounted at the SOAR telescope ($R\simeq2500, \lambda=3500-6160\,{\rm \AA}$). 

In addition to this we used spectra taken with the EMMI instrument ($R\simeq3400,\lambda=3900-4400\,{\rm \AA}$) mounted at the ESO-NTT and the UVES spectrograph ($R\simeq20000,\lambda=3300-6600\,{\rm \AA}$) mounted at the ESO-VLT in the course of the ESO Supernova Ia Progenitor Survey (SPY, Napiwotzki et al. \cite{napiwotzki03}). Data taken for studies of sdB binaries at high resolution (Edelmann et al. \cite{edelmann05}; Classen et al. \cite{classen11}) both with the FEROS spectrograph ($R\simeq48000,\lambda=3800-9200\,{\rm \AA}$) mounted at the ESO/MPG-2.2m telescope and with the Cross-Dispersed Echelle Spectrograph ($R\simeq60000,\lambda=3700-10000\,{\rm \AA}$) mounted at the McDonald observatory 2.7m telescope were used as well. Reduction was made either with the \texttt{MIDAS}, \texttt{IRAF} or \texttt{PAMELA} and \texttt{MOLLY}\footnote{http://www2.warwick.ac.uk/fac/sci/physics/research/astro/people\\/marsh/software} packages. 

\subsection{Photometry of HS\,2043+0615}

HS\,2043+0615 was extensively observed with the MEROPE camera at the Mercator telescope during the 2007 observing season. In total we used $R_C$-band photometry from 16 different nights, the first from April 22 and the last from November 15. The exposure time for these observations was 300\,s, and we collected in total 516 useful observations. Most of these runs spanned only a fraction of an orbit, and the photometric reduction was complicated because different setups and windows were used during the different runs, forcing us to use different reference stars for different runs. No standards were observed for these runs either, so to calibrate the photometry we made a catalogue of 22 stars in the field, starting by assigning R-band magnitudes from the NOMAD survey to each of the stars. These were then iteratively corrected until consistent values were achieved. The corrected magnitudes were then used to calibrate the differential photometry of HS\,2043+0615 to a common scale.

Recently, HS\,2043+0615 was reobserved with the three-channel MAIA camera on the Mercator telescope (Raskin et al. \cite{raskin13}).
MAIA splits the incoming light into three beams with dichroics to produce simultaneous photometry in a red, green, and UV channel, using three cameras equipped with large-format frame-transfer CCDs. We used an observing mode in which the $R$ and $G$ channels were read out every 120\,s and the $U$ channel only every second cycle to increase the signal-to-noise ratio. A run of 7\,h duration, almost a complete orbital cycle, was obtained on the night of September 3, 2013.

\begin{figure*}[t!]
\begin{center}
	\resizebox{8cm}{!}{\includegraphics{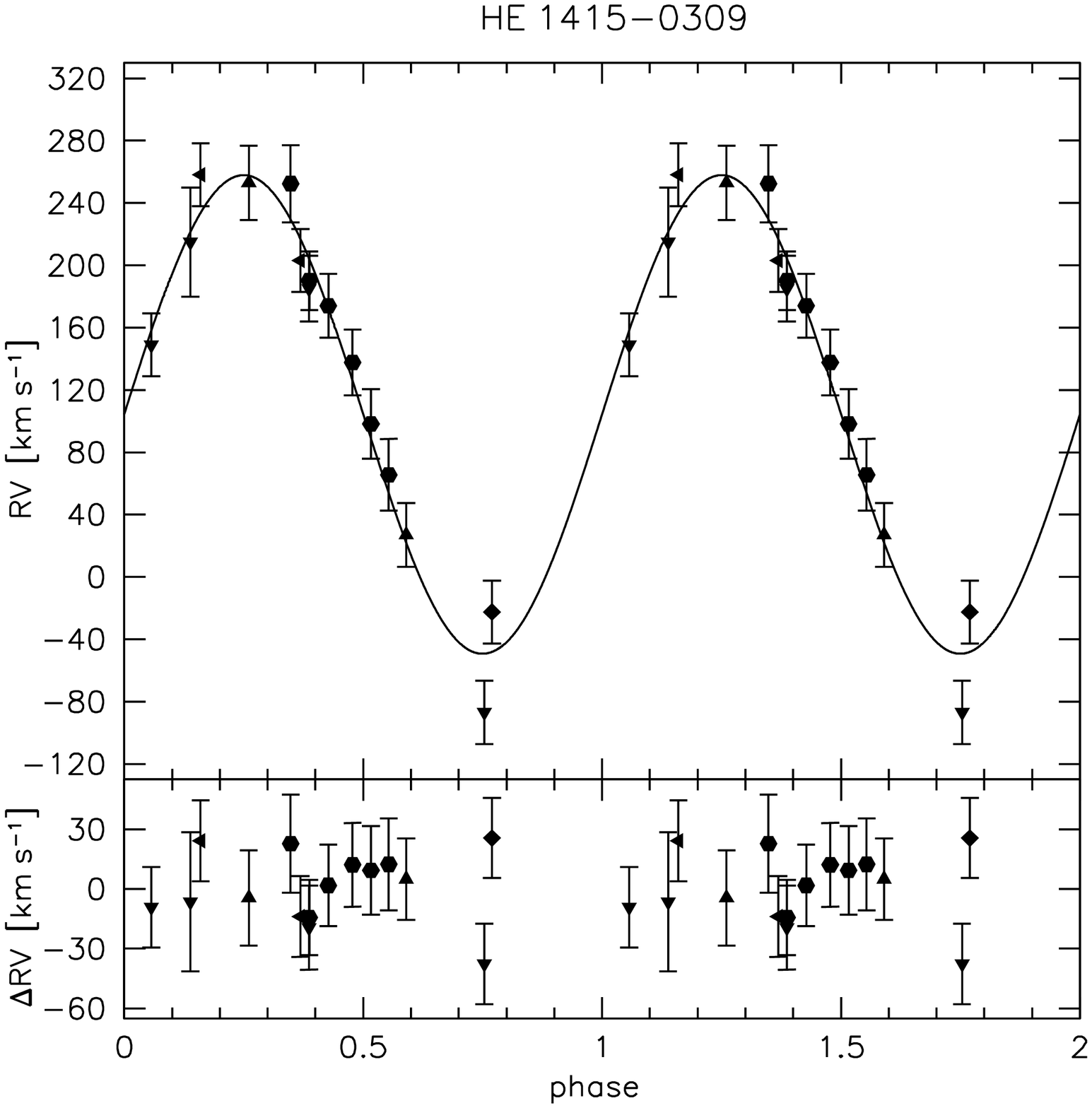}}
        \resizebox{8cm}{!}{\includegraphics{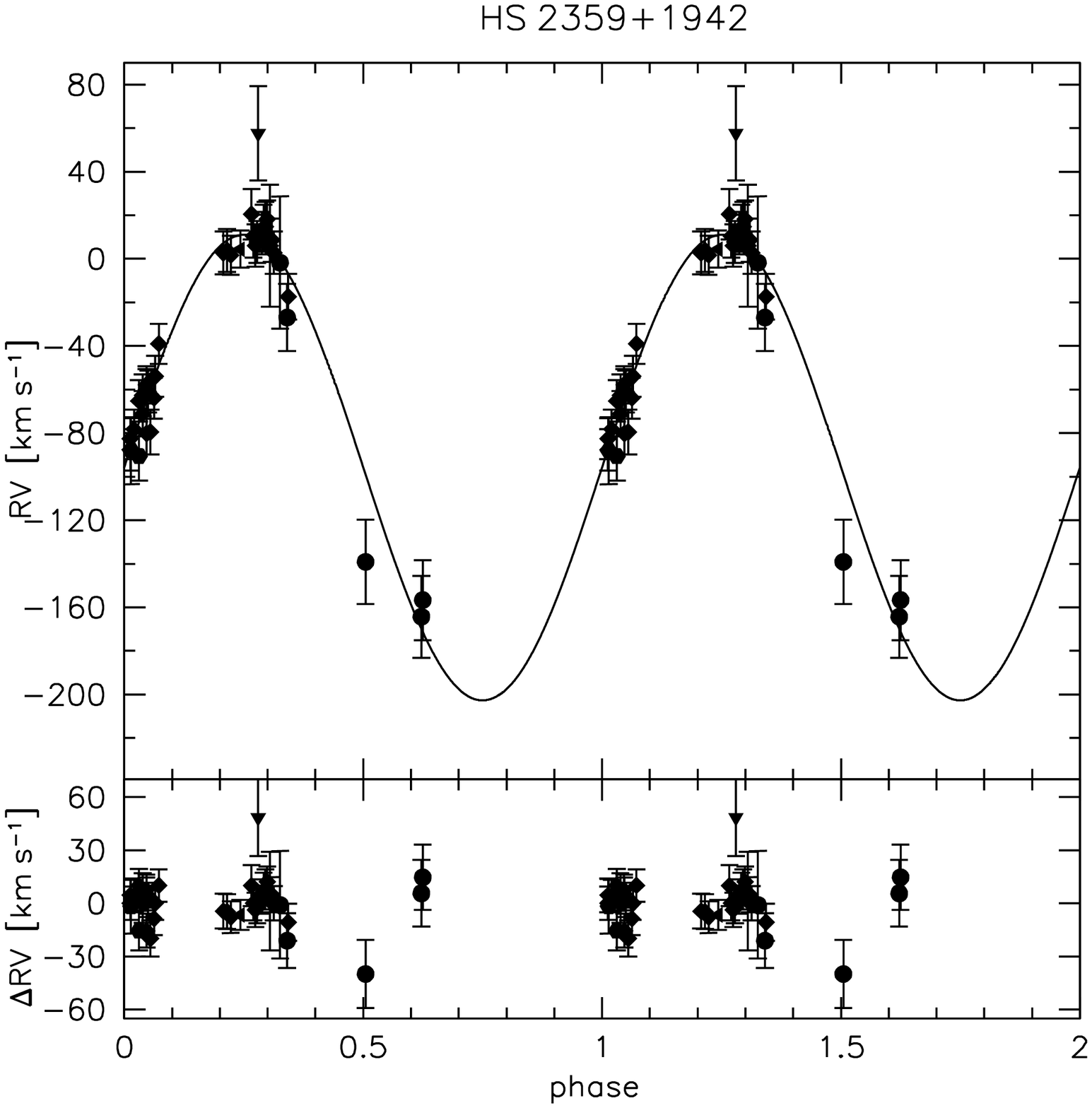}}
        \resizebox{8cm}{!}{\includegraphics{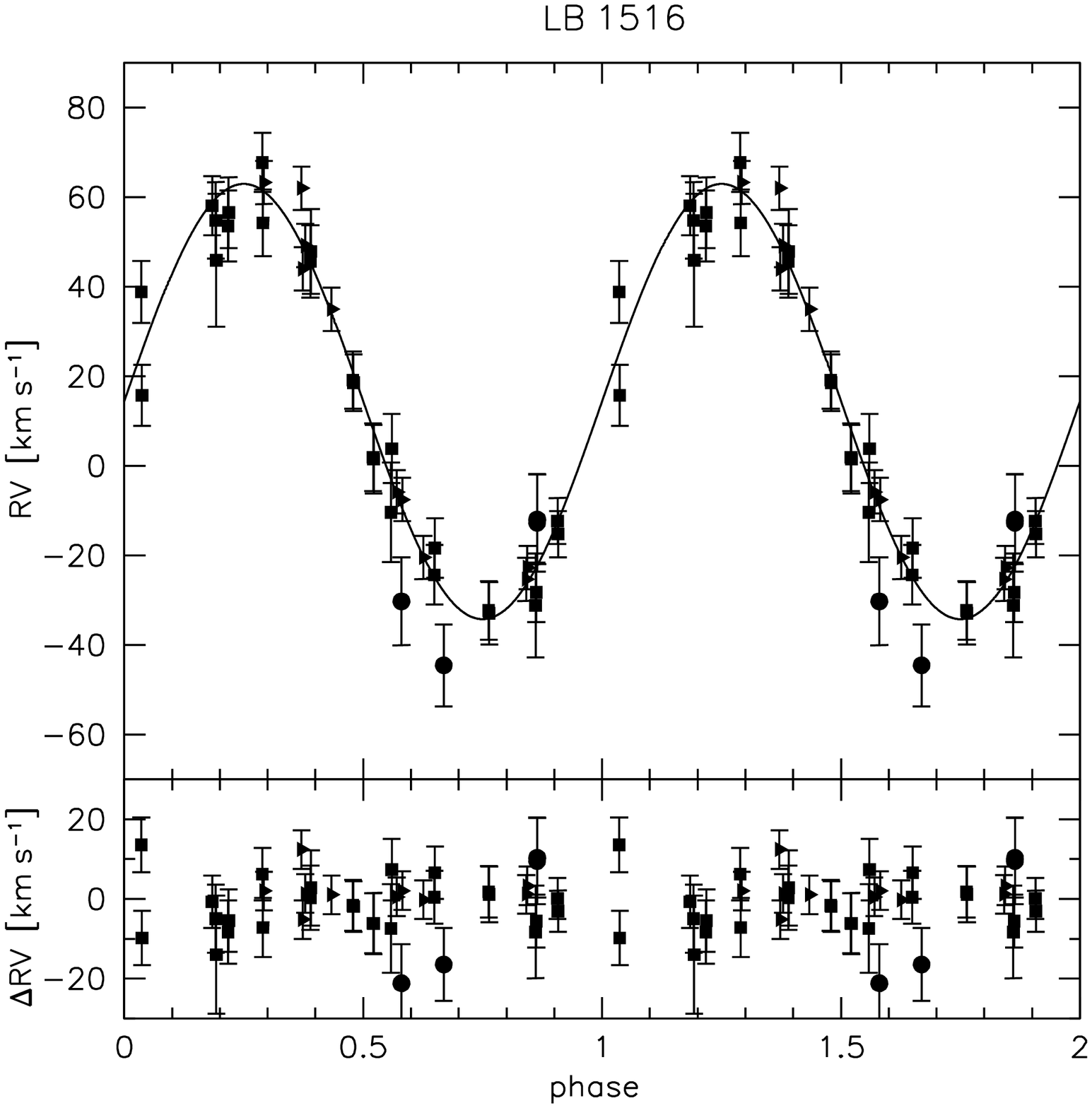}}
        \resizebox{8cm}{!}{\includegraphics{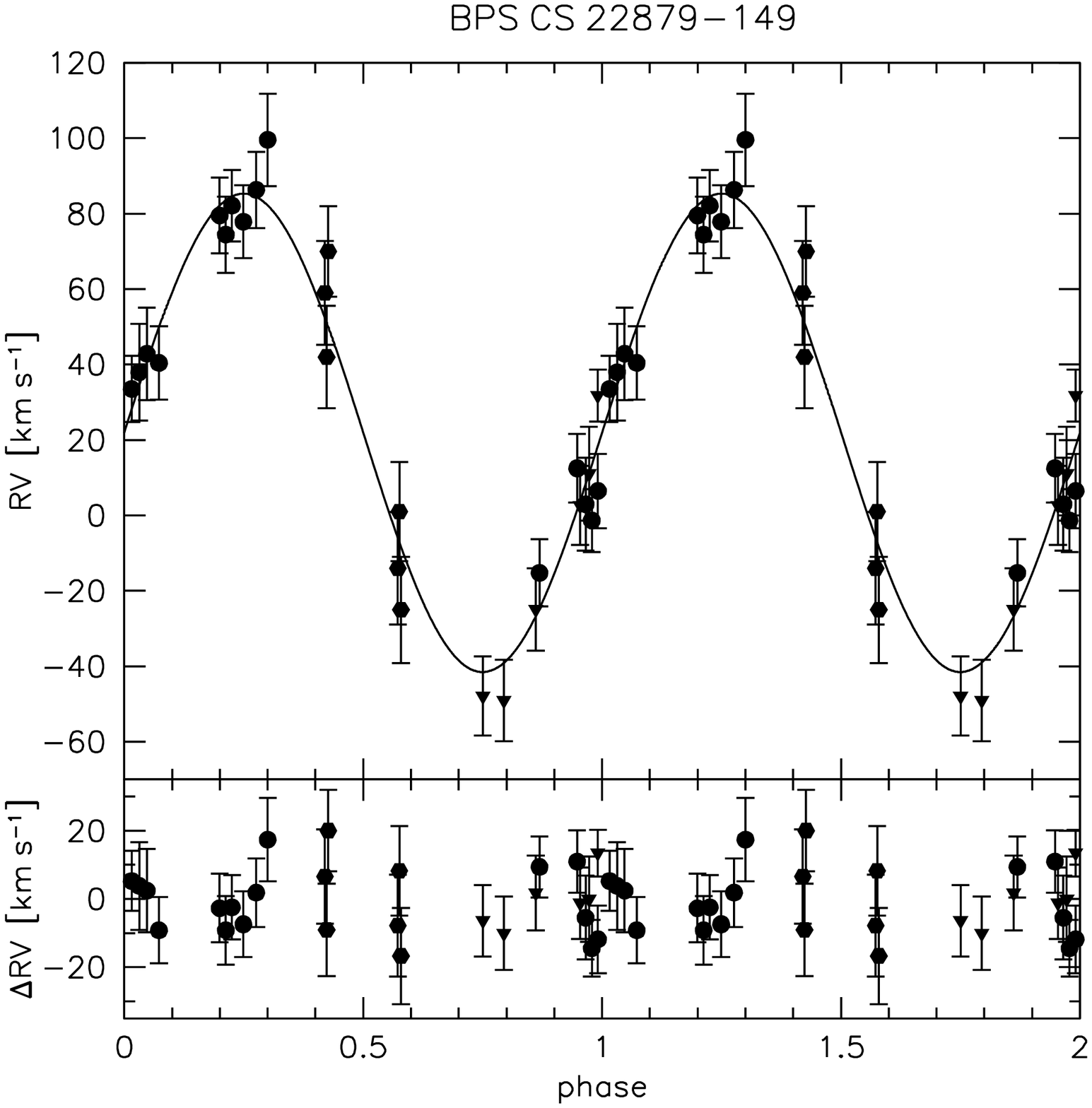}}
\end{center}
\caption{Radial velocity plotted against orbital phase. The RV data were phase folded with the most likely orbital periods. The residuals are plotted below. The RVs were measured from spectra obtained with CAHA-3.5m/TWIN (upward triangles), WHT/ISIS (diamonds), ESO-NTT/EMMI (downward triangles), ESO-VLT/UVES (triangles turned to the left), ESO-MPG2.2m/FEROS (triangles turned to the right), ESO-NTT/EFOSC2 (circles) and SOAR/Goodman (hexagons). RVs of LB\,1516 taken from Copperwheat et al. (\cite{copperwheat11}) are marked with rectangles. The RV data of BPS\,CS\,22879$-$149 was folded to the period alias at $0.478\,{\rm d}$.}
\label{rv1}
\end{figure*}

\begin{figure*}[t!]
\begin{center}
        \resizebox{6.0cm}{!}{\includegraphics{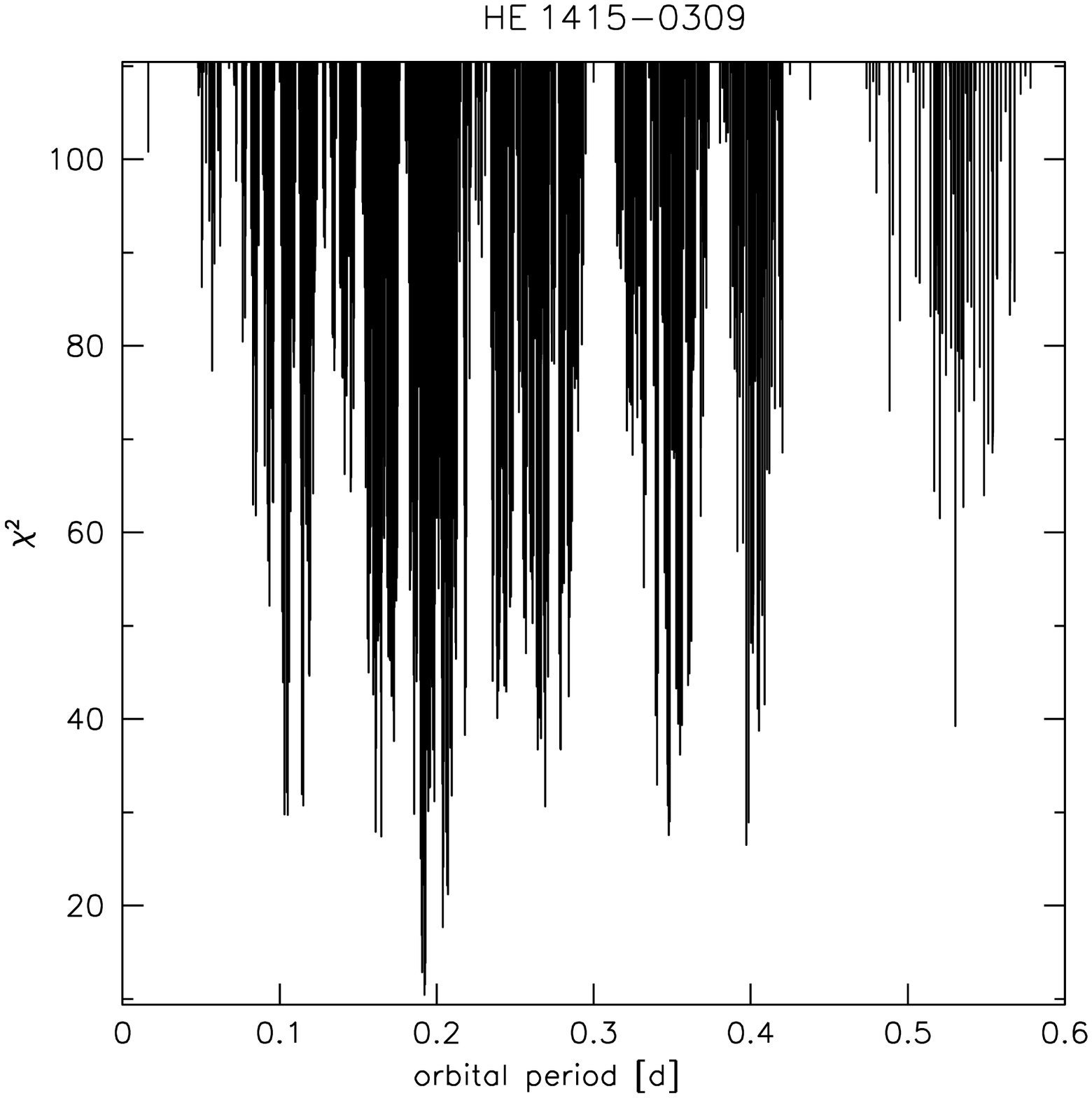}}
        \resizebox{6.0cm}{!}{\includegraphics{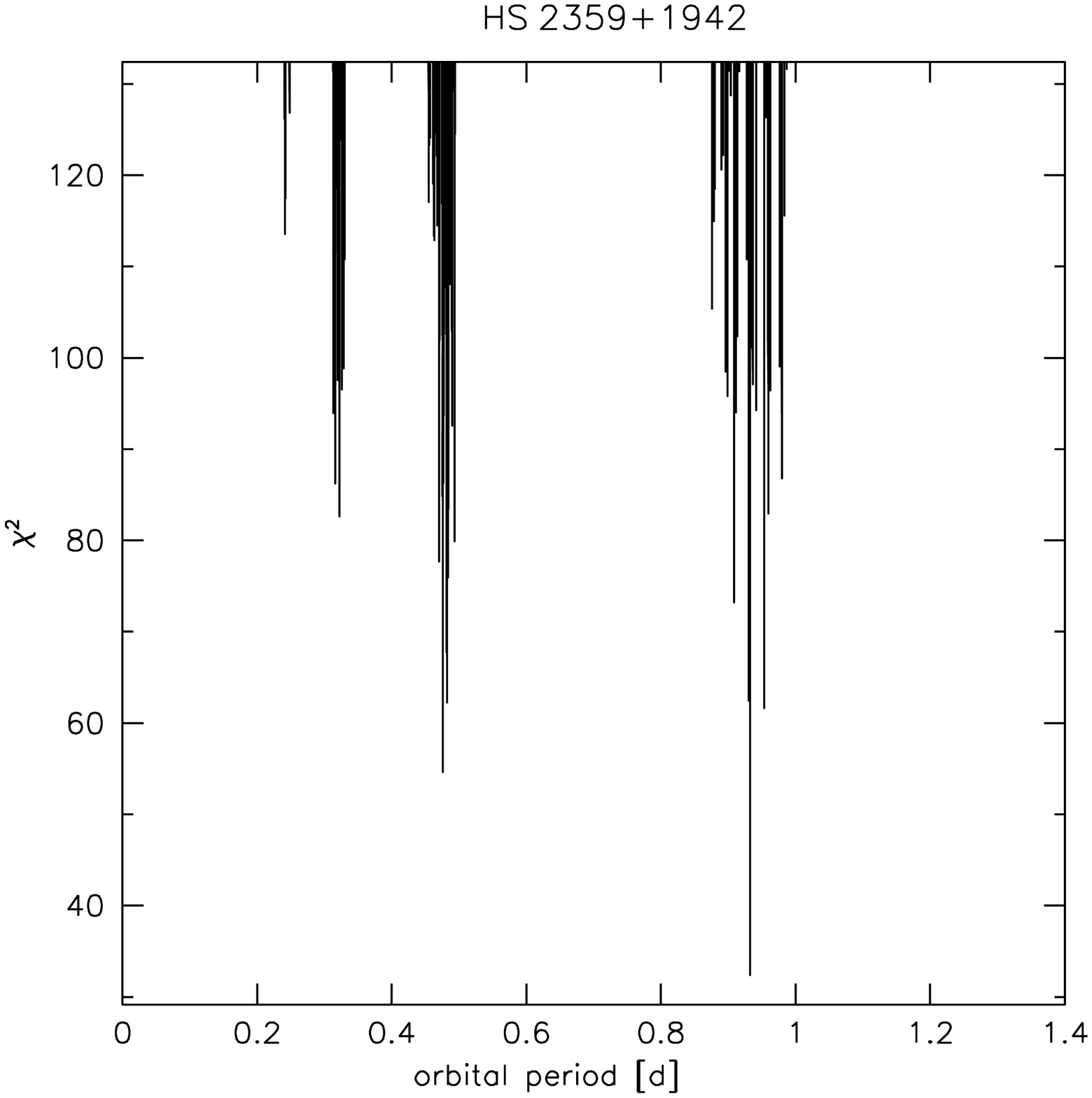}}
	\resizebox{6.0cm}{!}{\includegraphics{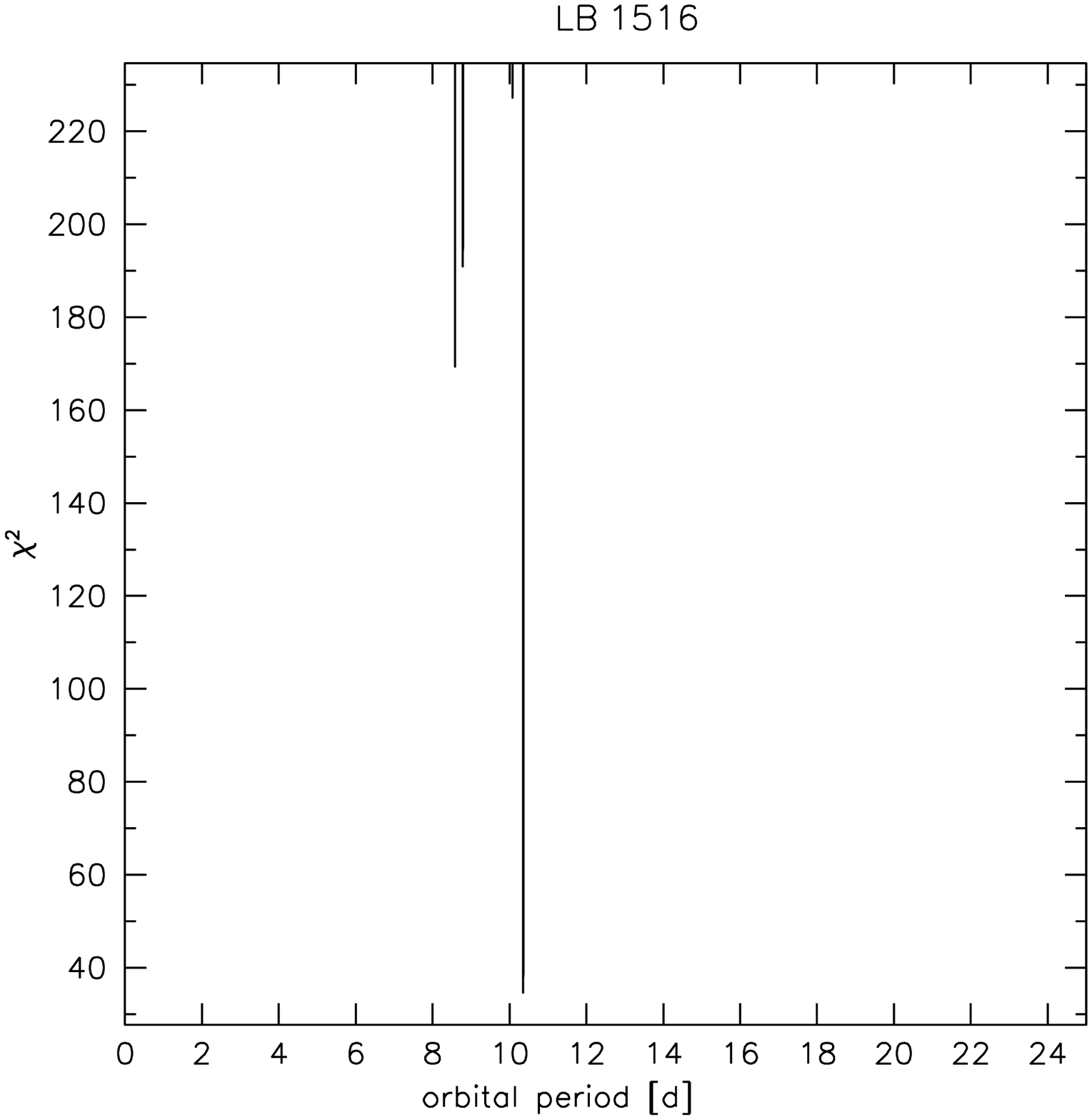}}
	\resizebox{6.0cm}{!}{\includegraphics{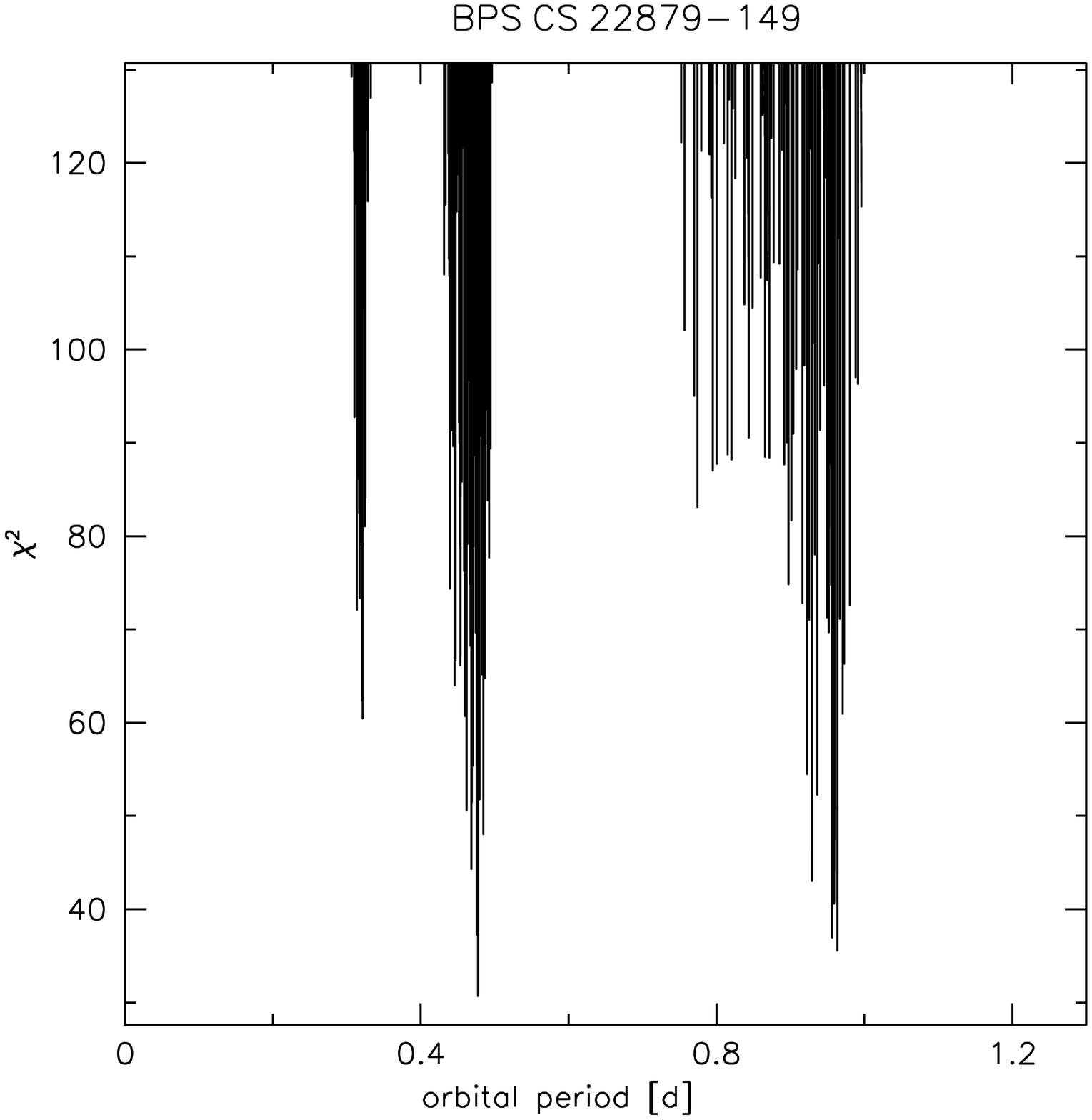}}
\end{center}
\caption{$\chi^{2}$ plotted against orbital period. The lowest peak corresponds to the most likely solution.}
\label{chi}
\end{figure*}

\section{Orbital and atmospheric parameters \label{s:orbit}}

The radial velocities were measured by fitting a set of mathematical functions to all suitable hydrogen Balmer as well as helium lines simultaneously, using $\chi^{2}$-minimization and the RV shift with respect to the rest wavelengths was measured (FITSB2, Napiwotzki et al. \cite{napiwotzki04b}). Gaussians were used to match the line cores, Lorentzians to match the line wings, and polynomials to match the continua. The RVs and formal $1\sigma$-errors are given in the appendix. 

For four binaries of our sample (HE\,1415$-$0309, HS\,2359+1942, LB\,1516, and BPS\,CS\,22879$-$149) the orbital parameters $T_{\rm 0}$, period $P$, system velocity $\gamma$, and RV-semiamplitude $K$ as well as their uncertainties and associated false-alarm probabilities ($p_{\rm false}[1\%]$, $p_{\rm false}[10\%]$) were determined as described in Geier et al. (\cite{geier11b}). To estimate the significance of the orbital solutions and the contributions of systematic effects to the error budget, we normalised the $\chi^{2}$ of the most probable solution by adding systematic errors $e_{\rm norm}$ in quadrature until the reduced $\chi^{2}$ reached $\simeq1.0$. The phased RV curves for the best solutions are given in Fig.~\ref{rv1}, the $\chi^{2}$-values plotted against orbital period in Fig.~\ref{chi}. The minimum in $\chi^{2}$ corresponds to the most likely solution. The adopted systematic errors and false-alarm probabilities are given in Table~\ref{tab:orbits}. The probabilities that the adopted orbital periods are correct to within $10\%$ range from $90\%$ to more than $99.99\%$. For BPS\,CS\,22879$-$149, no unique solution was found. The two possible solutions are discussed in Sect.~\ref{s:comp}.

\begin{figure*}[t!]
\begin{center} 
        \resizebox{8cm}{!}{\includegraphics{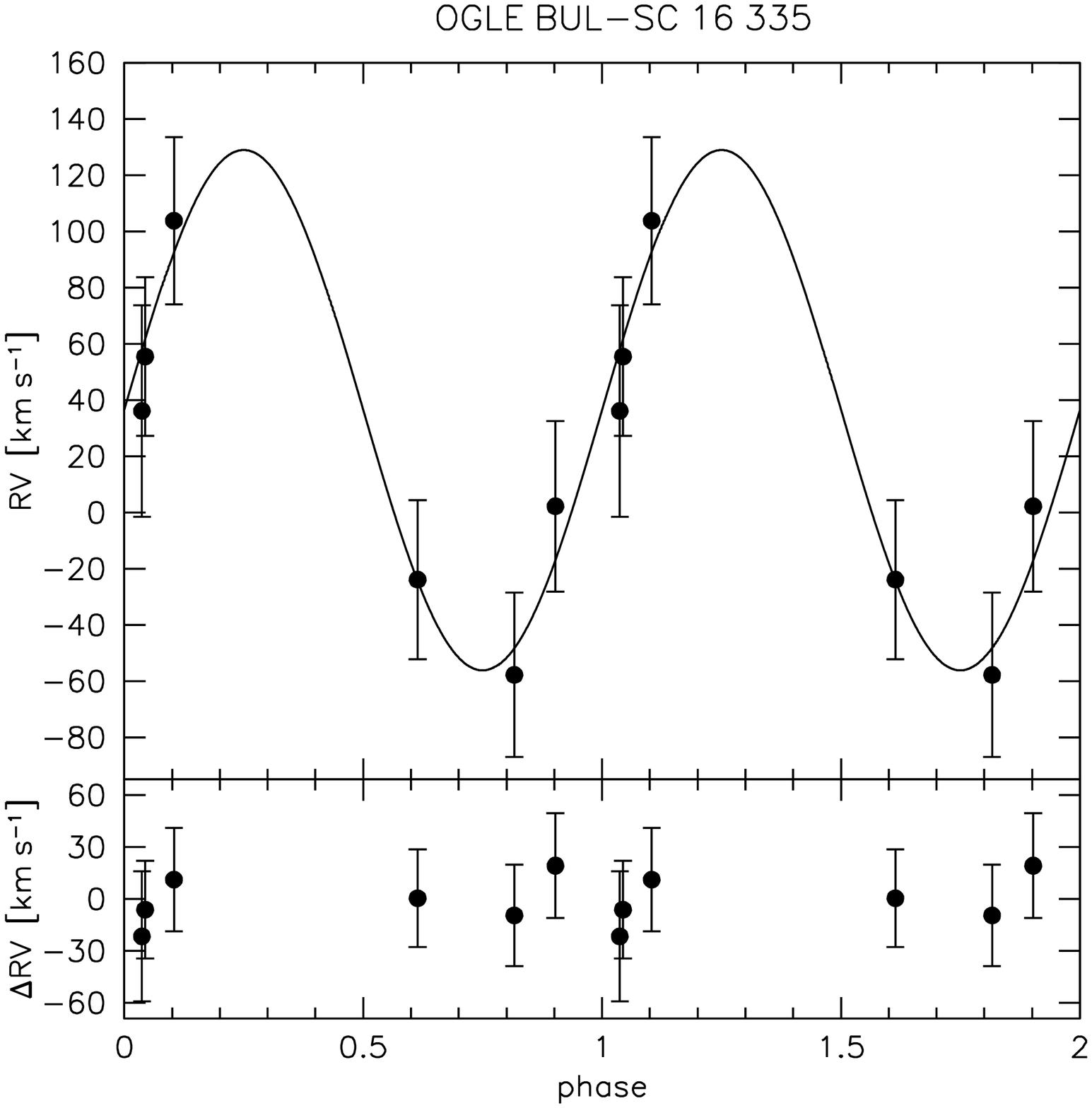}}
        \resizebox{8cm}{!}{\includegraphics{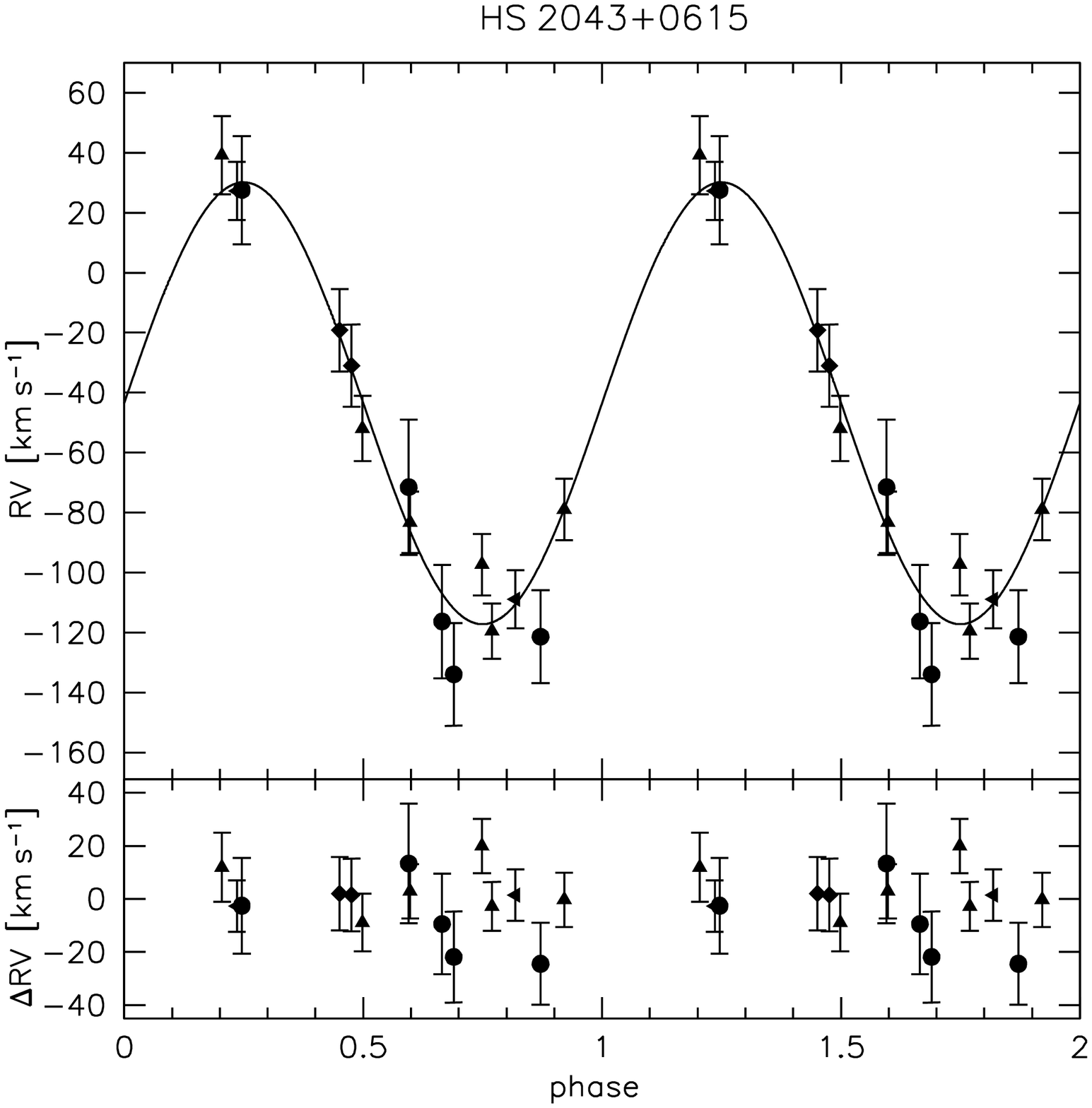}}
        \resizebox{8cm}{!}{\includegraphics{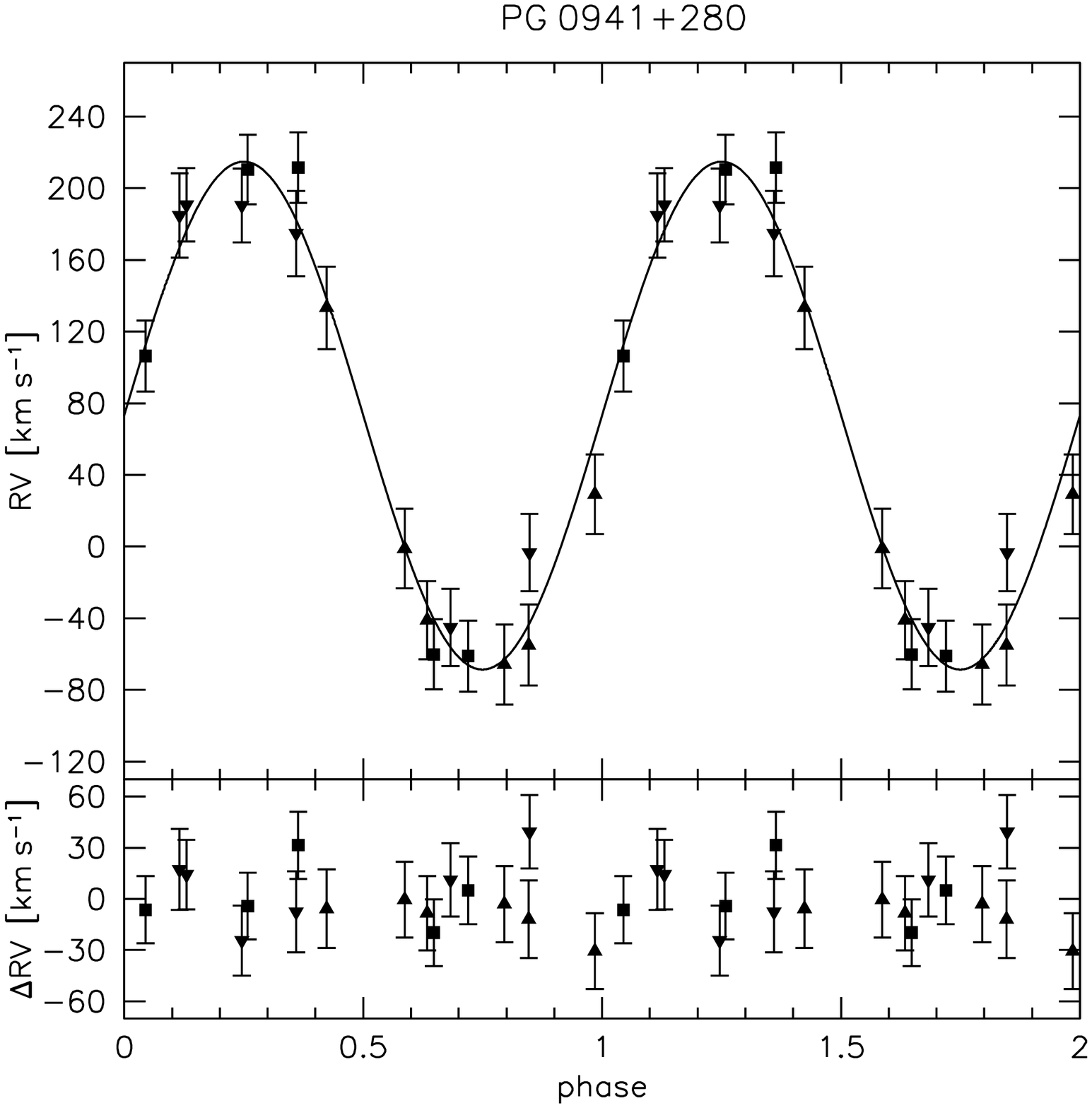}}
        \resizebox{8cm}{!}{\includegraphics{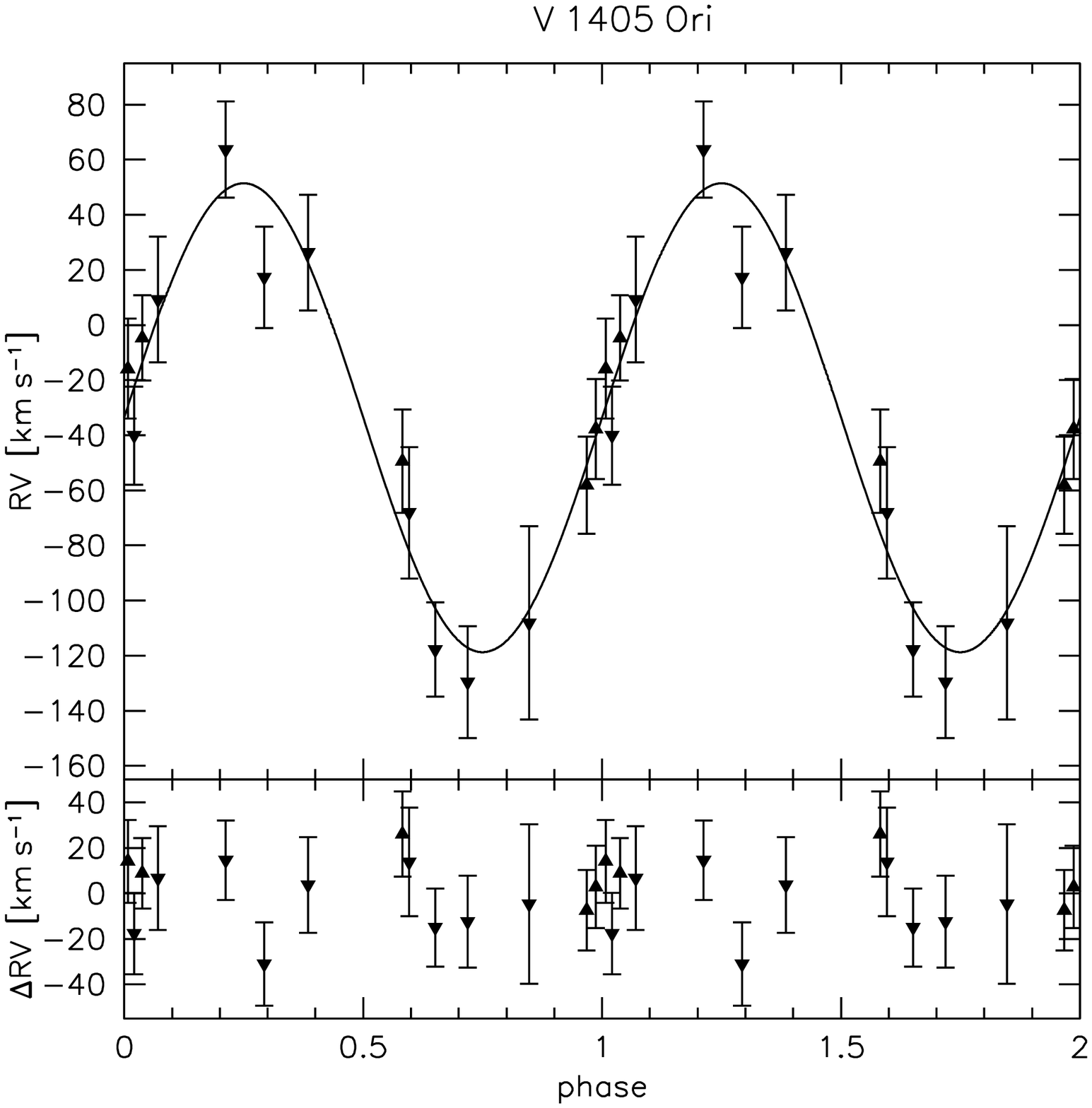}}
\end{center}
\caption{Phased radial velocity curves plotted twice for visualisation. The rectangles mark RVs measured from spectra obtained with McDonald-2.7m/Coude. The other symbols are the same as in Fig~\ref{rv1}.}
\label{rv2}
\end{figure*}

For OGLE\,BUL$-$SC16\,335 and V\,1405\,Ori the orbital period was independently determined from the variations seen in their light curves. These periods were kept fixed, but the other orbital parameters were determined in the way described above. The likely period of the eclipsing sdB+WD PG\,0941+280 was estimated from a light curve plotted in Green et al. (\cite{green04}) and compared with the period aliases derived from the RV measurements. The alias closest to the estimate from the light curve was identified as solution. A similar approach was chosen for the reflection effect binary HS\,2043+0615, for which the most likely period of the light curve was compared with the corresponding alias periods of the radial velocity curve. The phased radial velocity curves of those binaries are shown in Fig.~\ref{rv2}.

The atmospheric parameters effective temperature $T_{\rm eff}$, surface gravity $\log{g}$ and helium abundance $\log{y}$ of PG\,0941+280, V\,1405\,Ori and OGLE\,BUL$-$SC16\,335 were determined as described in Geier et al. (\cite{geier11a}) by fitting model atmospheres with local thermodynamic equilibrium and supersolar metallicity (O'Toole \& Heber \cite{otoole06}) to the hydrogen and helium lines of a coadded spectrum. For PG\,0941+280 and V\,1405\,Ori we used a TWIN spectrum, whereas an EFOSC2 spectrum was used for  OGLE\,BUL$-$SC16\,335. 

\begin{table*}[t!]
\caption{Derived orbital solutions.} 
\label{tab:orbits}
\begin{center}
\begin{tabular}{lllrrlll} 
\hline\hline
\\[-3mm]
Object & $T_{0}$ & P &  $\gamma$ & K & $e_{\rm norm}$ & $\log{p_{\rm false}}[10\%]$ & $\log{p_{\rm false}}[1\%]$ \\
 & [HJD$-$2\,450\,000] & [d] & [${\rm km\,s^{-1}}$] & [${\rm km\,s^{-1}}$] & [${\rm km\,s^{-1}}$] & &  \\ 
\hline
\\[-3mm]
HE\,1415$-$0309 & $5240.909\pm0.002$ & $0.192\pm0.004$   & $104.7\pm9.5$ & $152.4\pm11.2$ & $18.6$ & $-1.0$ & $-0.4$ \\
HS\,2359+1942 & $6279.221\pm0.007$ & $0.93261\pm0.00005$ & $-96.1\pm6.0$ & $107.4\pm6.8$ & $8.3$ & $-1.2$ & $-1.1$ \\
LB\,1516 & $5495.73\pm0.05$  & $10.3598\pm0.0005$     & $14.3\pm1.1$  & $48.6\pm1.4$ & $4.4$  & $<-4.0$ & $<-4.0$ \\
\hline \\[-3mm]
BPS\,CS\,22879$-$149 & $5413.102$ & $0.478$ & $21.9\pm2.5$ & $63.5\pm2.8$ & $5.4$ & ...  & ... \\
                     & $5412.448$ & $0.964$ & $-25.5\pm5.3$ & $121.7\pm6.3$ & $6.2$ & ... & ... \\
\hline \\[-3mm]
OGLE\,BUL$-$SC16\,335  & $4758.614$         & $0.122$           & $36.4\pm19.6$ & $92.5\pm26.2$  & $25.1$ & ... & ... \\
HS\,2043+0615   & $4254.610\pm0.003$ & $0.3015\pm0.0003$ & $-43.5\pm3.4$ & $73.7\pm4.3$   & $8.3$  & ... & ... \\
PG\,0941+280    & $4476.185$         & $0.311$           & $73.0\pm4.9$  & $141.7\pm6.3$  & $19.4$  & ... & ... \\
V\,1405\,Ori    & $4477.362$         & $0.398$           & $-33.6\pm5.5$ & $85.1\pm8.6$   & $15.2$ & ... & ... \\  
\hline \\[-3mm]
\end{tabular}
\tablefoot{The systematic error adopted to normalise the reduced $\chi^{2}$ ($e_{\rm norm}$) is given for each case. The probabilities for the orbital period to deviate from our best solution by more than $10\%$ ($p_{\rm false}[10\%]$) or $1\%$ ($p_{\rm false}[1\%]$) are given in the last columns. The last four lines show the binaries, where the orbital period has been determined from photometry.}
\end{center}
\end{table*}

\section{Nature of the unseen companions}

\subsection{Methods to constrain the nature of the companion}

All spectroscopic binaries in our sample are single-lined and their binary mass functions can be determined from

\begin{equation}
 \label{equation-mass-function}
 f_{\rm m} = \frac{M_{\rm comp}^3 \sin^3i}{(M_{\rm comp} +
   M_{\rm sdB})^2} = \frac{P K^3}{2 \pi G}
\end{equation}

The RV semi-amplitude and the orbital period can be derived from the RV curve, but the sdB mass $M_{\rm sdB}$, the companion mass $M_{\rm comp}$ and the inclination angle $i$ remain free parameters. Adopting the canonical sdB mass $M_{\rm sdB}=0.47\,{\rm M_{\odot}}$ (see discussion in Fontaine et al. \cite{fontaine12}) and $i<90^{\rm \circ}$, we derive a lower limit for the companion masses (see Table\,\ref{rvmasses}). For mini\-mum companion masses lower than $\sim0.45\,M_{\rm \odot}$ the companion may be a late-type main sequence star or a compact object such as a WD. Main sequence stars in this mass range are outshone by the sdBs and are not visible in optical spectra (Lisker et al. \cite{lisker05}). If on the other hand the minimum companion mass exceeds $0.45\,M_{\rm \odot}$, spectral features of a main sequence companion become visible in the optical. The non-detection of such features therefore allows us to exclude a main sequence star. 

Indicative features in the light curves constrain the nature of the companions further in some cases. A sinusoidal variation with orbital period originates from  the irradiation of a cool companion by the hot subdwarf primary. The projected area of the companion's heated hemisphere changes while it orbits the primary. The detection of this so-called reflection effect indicates a cool companion with a size similar to the hot subdwarf primary, either a low-mass main sequence star of spectral type M or a substellar object such as a brown dwarf. If eclipses are present as well, the inclination angle can be measured and the mass of the companion can be constrained. Such eclipsing sdB binaries with reflection effect are also known as HW\,Vir-type binaries. 

The lack of variations in the light curve, on the other hand, can be used to exclude a cool companion, when the orbital period of the binary is sufficiently short. In this case a reflection effect would be easily detectable and a non-detection implies that the companion must be a compact object. The detection of the very shallow eclipses from a compact WD companion also allows us to constrain its mass. Smaller variations indicative of a massive compact companion, which are caused by the ellipsoidal deformation and the relativistic Doppler beaming of the sdB primary, can only be detected from the ground in the most extreme cases (e.g. Geier et al. \cite{geier07,geier13}). However, using high-precision space-based photometry, these variations can be detected and used to constrain the binary parameters (Geier et al. \cite{geier08}; Bloemen et al. \cite{bloemen11}; Telting et al. \cite{telting12}).

\begin{table}[t!]
\caption{Derived masses and most probable nature of the companions.} 
\label{rvmasses}
\begin{center}
\begin{tabular}{llll}
\hline\hline
\\[-3mm] 
Object & $f(M)$ & $M_{\rm 2}$ & Companion \\
 & [$M_{\rm \odot}$] & [$M_{\rm \odot}$] &  \\ 
\hline
\\[-3mm]
HE\,1415$-$0309  & $0.07$ & $>0.37$ & WD \\
PG\,0941+280     & $0.092$ & $0.42\pm0.03$ & WD \\
HS\,2359+1942    & $0.12$ & $>0.47$ & WD \\
LB\,1516         & $0.12$ & $>0.48$ & WD \\ 
\hline \\[-3mm]
OGLE\,BUL$-$SC16\,335  & $0.01$  & $0.16\pm0.05$ & MS \\
HS\,2043+0615        & $0.013$ & $0.18-0.34$ & MS \\
V\,1405\,Ori         & $0.034$ & $>0.26$ & MS \\
\hline \\[-3mm]
BPS\,CS\,22879$-$149 & $0.013$ & $>0.18$ & MS/WD  \\
                     & $0.17$  & $>0.57$ & WD \\
\hline \\[-3mm]
\end{tabular}
\end{center}
\end{table}

\subsection{White dwarf companions}

{\bf HE\,1415$-$0309} has been identified as a single-lined sdB star in the course of the SPY project (Lisker et al. \cite{lisker05}). A significant shift in radial velocity ($\sim130\,{\rm km\,s^{-1}}$), indicating a close binary, has been measured from two UVES spectra (Napiwotzki priv. comm.). The minimum mass of the companion is too low ($0.37\,M_{\rm \odot}$) to exclude a main sequence star. However, a light curve of this star ($\sim1\,{\rm hr}$) was taken with the Nordic Optical Telescope on La Palma to search for pulsations and no variations have been reported (\O stensen et al. \cite{oestensen10b}). Due to the short orbital period of only $4.6\,{\rm hr}$, a reflection effect would have been easily detectable. We therefore conclude that the unseen companion of HE\,1415$-$0309 must be a compact object, most likely a WD. 

{\bf HS\,2359+1942} (PG\,2359+197) was drawn from the SPY sample and ana\-lysed by Lisker et al. (\cite{lisker05}). We detected an RV shift of an EMMI spectrum with respect to the survey spectrum taken with UVES. The minimum companion mass is $0.47\,M_{\rm \odot}$, similar to the adopted mass of the sdB itself. Since no spectral features of a cool MS companion have been found, we conclude that the companion must be a compact object, presumably a WD. 

{\bf LB\,1516} (EC\,22590$-$4819) has been discovered to be an sdB binary with a period of a few days by Edelmann et al. (\cite{edelmann05}), but no unambiguous solutions was found. Koen et al. (\cite{koen10}) identified the sdB to be a g-mode pulsator. Subsequently, Copperwheat et al. (\cite{copperwheat11}) obtained an orbital solution of this system ($P=10.3592$, $K=46.8\pm1.8\,{\rm km\,s^{-1}}$). We combined the RV measurements from Edelmann et al. (\cite{edelmann05}) and Copperwheat et al. (\cite{copperwheat11}) with additional RVs measured from FEROS spectra and our new measurements to obtain a more accurate solution. The rather long period of $10.3958\,{\rm d}$ leads to a minimum companion mass of $0.48\,M_{\rm \odot}$. Since no spectral features of the companion are detectable, the companion is likely to be a WD.

{\bf PG\,0941+280} (HX\,Leo) has been identified as an sdB star by Saffer et al. (\cite{saffer94}). The effective temperature $T_{\rm eff}=29400\pm500\,{\rm K}$, surface gravity $\log{g}=5.43\pm0.05$ and helium abundance $\log{y}=-3.0\pm0.1$ are consistent with the results ($T_{\rm eff}=29000\pm1000\,{\rm K}$, $\log{g}=5.58\pm0.15$, $\log{y}=-3.0$) of Saffer et al. (\cite{saffer94}), who used pure hydrogen models. 

Green et al. (\cite{green04}) detected shallow eclipses of an earth-sized WD companion in the light curve. The orbital period estimated from these eclipses is around $0.3\,{\rm d}$. We adopted the period alias of our RV analysis closest to this result as the most likely orbital period. The derived mass of the companion assuming $\sin{i}=1$ is $0.42\pm0.03\,M_{\rm \odot}$. 

\subsection{M-dwarf companions}

{\bf OGLE\,BUL$-$SC16\,335} was identified as an HW\,Vir system by Polubek et al. (\cite{polubek07}). Since this analysis was based on photometry alone, the sdB nature of the primary could not be proven unambiguously. We constrained the atmospheric parameters of OGLE\,BUL$-$SC16\,335 by fitting model spectra. Due to the limited wavelength range we were only able to use H$_{\rm \beta}$ and the two He\,{\sc i} lines at $4472\,{\rm \AA}$ and $4922\,{\rm \AA}$. However, within the uncertainties the resulting effective temperature $T_{\rm eff}=31500\pm1800\,{\rm K}$, surface gravity $\log{g}=5.7\pm0.2$, and helium abundance $\log{y}=-1.8\pm0.1$ are perfectly consistent with an sdB primary. 

Adopting the orbital period derived from the light curve by Polubek et al. (\cite{polubek07}), we determined the RV semiamplitude and the mass of the companion ($0.16\pm0.05\,M_{\rm \odot}$). A substellar companion can be excluded and the companion is a low-mass M-dwarf. 

\begin{figure*}[t!]
	\resizebox{16cm}{!}{\includegraphics[angle=0]{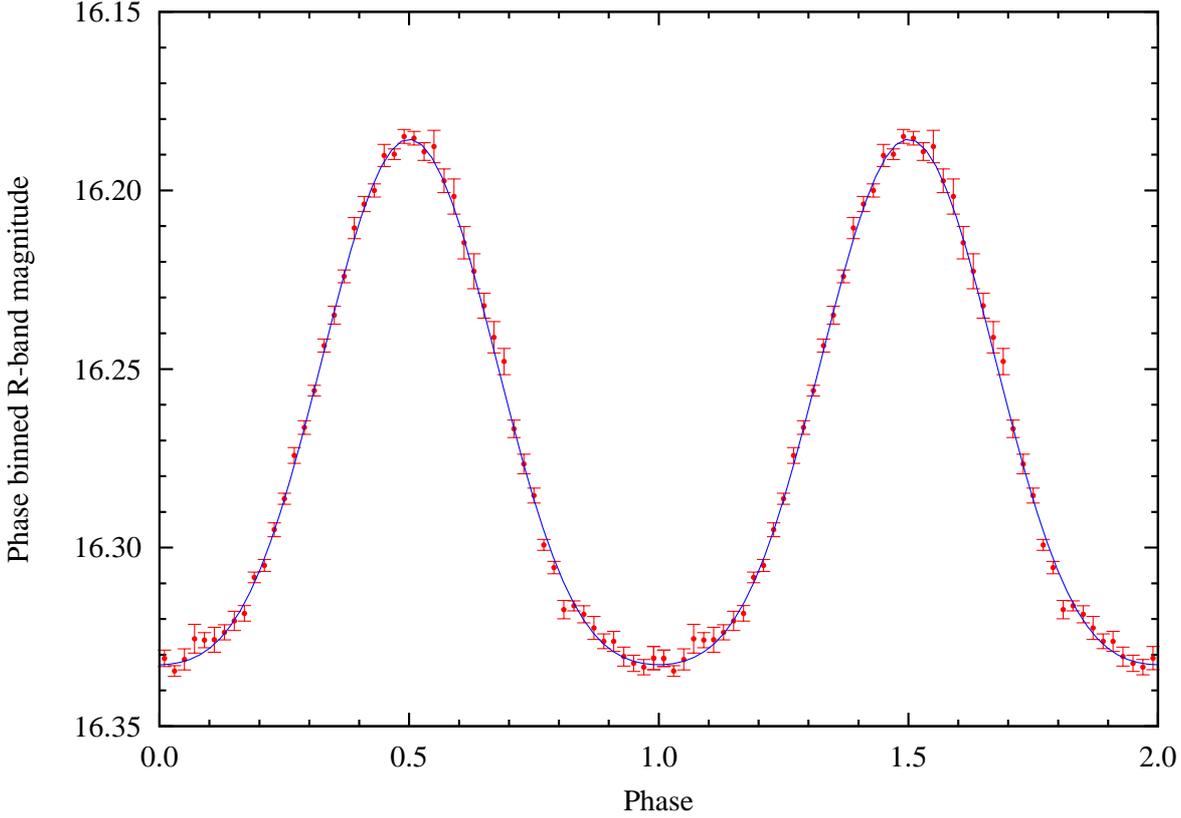}}
	\caption{Phase-binned light curve of HS\,2043+0615. The 516 MEROPE data points from 2007 were phase-folded on P\,=\,0.30156 day using 50 bins, and are plotted twice to better visualise the difference in width of the peaks and throughs. The error bars indicate the rms for each phase bin.}
\label{fig:merope_lc}
\end{figure*}

\begin{figure}[t!]
	\resizebox{\hsize}{!}{\includegraphics[angle=0]{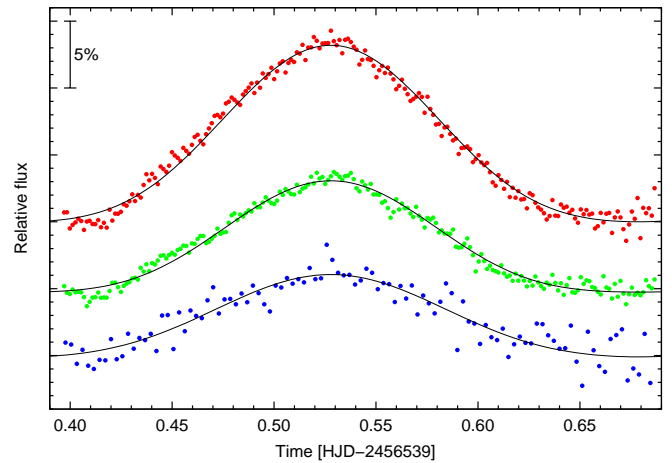}}
	\caption{Multiband light curve of HS\,2043+0615 taken with the MAIA camera (R, G, U-bands from top to bottom).}
\label{fig:maia_lc}
\end{figure}

{\bf HS\,2043+0615} was again drawn from the SPY sample and analysed by Lisker et al. (\cite{lisker05}). A shift in radial velocity ($\sim135\,{\rm km\,s^{-1}}$) has been measured from two UVES spectra (Napiwotzki priv. comm.). As mentioned by \O stensen et al. (\cite{oestensen10b}), it was observed on two consecutive nights in June 2005 with the NOT, and found to have a strong variability with a period of several hours, presumably due to a reflection effect. An extensive photometric follow-up has then been conducted with the Mercator telescope.

To determine the ephemeris we phase-folded the seven months of MEROPE photometry on different trial periods and selected that with the lowest variance. The resulting light curve after folding into 50 phase bins is shown in Fig.~\ref{fig:merope_lc}. There are no significant competing aliases in the periodogram. As there are no sharp eclipses that can be used to accurately phase observations at different epochs, the error on the period is quite large. We estimate that we can phase our data to a precision of 1/10 of a cycle, and since our useful observations span 206.8\,d\,$\approx$\,686 cycles, the phase error would be $\sim$0.3/10/686. We thus state the ephemeris as $T_{\rm 0}=2454213.70\pm0.03$ and $P=0.30156\pm0.00005\,{\rm d}$, perfectly consistent with the corresponding alias of the RV periodogram. The MAIA multiband light curves are plotted in Fig.~\ref{fig:maia_lc}. We fitted the light curves with a pair of phase-locked cosine functions as in \O stensen et al. (\cite{oestensen13}), eq.~1, and these are plotted with solid lines in Fig.~\ref{fig:maia_lc}. The semi-amplitudes for the orbital period, $A$, and for the first harmonic, $B$, are given in Table~\ref{tbl:hs2043_amp}.

\begin{figure}[t!]
	\resizebox{\hsize}{!}{\includegraphics[angle=0]{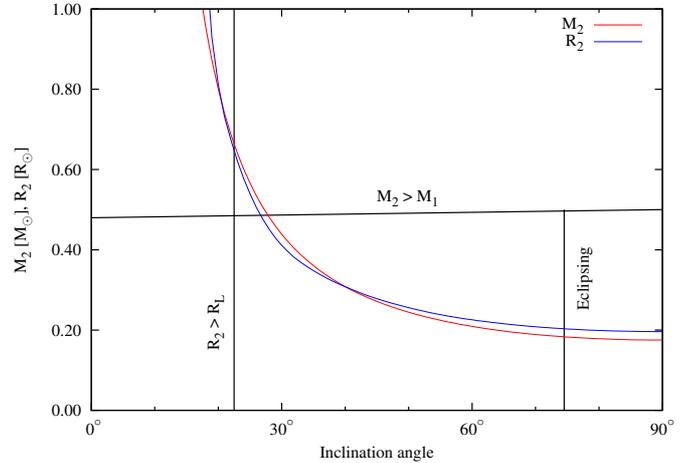}}
	\caption{Mass and radius of the M-dwarf companion to HS\,2043+0615 as a function of inclination angle, as indicated by the mass function and the mass-radius relation for M-dwarfs. The box limited by the three solid lines marks the possible parameter range. The companion cannot fill its Roche lobe $R_{\rm L}$, the binary is not eclipsing, and the companion mass cannot not be higher than the mass of the subdwarf, because it would then be visible in the optical spectrum (for details see \O stensen et al. \cite{oestensen13}).}
\label{fig:hs2043_mass}
\end{figure}

From the orbital solution we derive a minimum companion mass of $0.17\,M_{\rm \odot}$ consistent with an M-dwarf companion. Following the simple modelling approach described in \O stensen et al. (\cite{oestensen13}) and adopting the atmospheric parameters of HS\,2043+0615 given in Lisker et al. (\cite{lisker05}) as well as the theoretical mass-radius relation for M-dwarfs from Baraffe et al. (\cite{baraffe98}), we constrain the likely range of orbital inclinations to $30^{\rm \circ}<i<75^{\rm \circ}$ and the companion mass range to $0.18\,M_{\rm \odot}<M_{\rm comp}<0.34\,M_{\rm \odot}$ (see Fig.~\ref{fig:hs2043_mass}). 

\begin{table}
\caption{Photometric amplitudes from three-channel photometry.}
\label{tbl:hs2043_amp}
\begin{center}
\begin{tabular}{cll}
\hline\hline \noalign{\smallskip}
Band & A & B \\
\noalign{\smallskip} \hline \noalign{\smallskip}
$R$ & 0.0660(6) & 0.0119(6) \\
$G$ & 0.0416(6) & 0.0084(6) \\
$U$ & 0.0307(14) & 0.0041(14) \\
\noalign{\smallskip} \hline
\end{tabular}
\end{center}
\end{table}

{\bf V\,1405\,Ori} was discovered to be a short-period sdB pulsator (Koen et al. \cite{koen99}) with a reflection effect (Reed et al. \cite{reed10}). We phased our RVs to the orbital period determined from the light curve ($0.398\,{\rm d}$) and derived a minimum mass of $0.26\,M_{\rm \odot}$ for the M-dwarf companion. The effective temperature $T_{\rm eff}=35100\pm800\,{\rm K}$, surface gravity $\log{g}=5.66\pm0.11$, and helium abundance $\log{y}=-2.5\pm0.2$ are quite typical for short-period sdB pulsators of V\,361\,Hya type (see \O stensen \cite{oestensen10} and references therein). 

\begin{figure}[t!]
	\resizebox{\hsize}{!}{\includegraphics{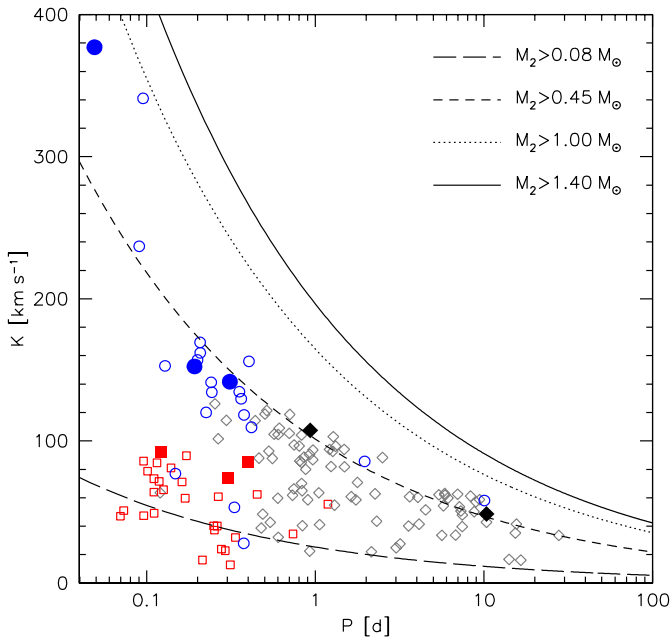}}
	\caption{RV semiamplitudes of all known sdB binaries with spectroscopic solutions plotted against their orbital periods (Kupfer et al. \cite{kupfer13}). Circles mark sdB binaries with compact companions, squares systems with M-dwarf or substellar companions, and diamonds binaries where the nature of the companion cannot be further constrained by photometry. The lines mark the regions to the right where the minimum companion masses derived from the binary mass function (assuming $0.47\,{\rm M_{\odot}}$ for the sdBs) exceed certain values. The binaries from the MUCHFUSS supplementary programme including CD$-$30$^\circ$11223 (Geier et al. \cite{geier13}) are marked with filled symbols, binaries taken from the literature with open symbols.}
\label{periodK}
\end{figure}

\subsection{Unconstrained companion type \label{s:comp}}

{\bf BPS\,CS\,22879$-$149} was identified as an sdB star by Beers et al. (\cite{beers92}) and chosen as a bright backup target for the southern sky. Since no unique orbital solution could be found, the minimum mass of the companion is either constrained to $0.18\,M_{\rm \odot}$ or to $0.57\,M_{\rm \odot}$. While in the latter case a WD companion would be most likely, a compact object of low mass is possible as well as an M dwarf in the former case. Comparing the two orbital solutions with the sample of known sdB binaries, the long-period solution appears to be more likely, because the number of known sdB binaries with such orbital parameters is higher than the number of binaries with the short-period parameters (see Fig.~\ref{periodK}). However, selection effects also favour the detection of higher RV-shifts. We conclude that the nature of the companion cannot be firmly constrained at this point.

\section{Discussion}

We derived orbital solutions of eight close hot subdwarf binaries and constrained the most likely nature of the unseen companions in all cases but one, using additional information derived from photometry and spectroscopy. These binaries cover the full parameter range of the known close binary sdB population (see Fig.~\ref{periodK}). Their companion types are also consistent with the apparent split between M-dwarf and substellar companions on one hand and WD companions on the other hand, especially at short orbital periods $<0.3\,{\rm d}$. Furthermore, our sample contains some peculiar binaries that deserve a more detailed analysis in the future.

PG\,0941$+$280 is one of only five sdB+WD binaries, where the shallow eclipses of the white dwarf have been detected in the light curves (the others are KPD\,0422+5421, Orosz \& Wade \cite{orosz99}; PG\,2345+318, Green et al. \cite{green04}; KPD\,1946+4340, Bloemen et al. \cite{bloemen11}; CD$-$30$^\circ$11223, Geier et al. \cite{geier13}). Its mass of $0.42\,M_{\rm \odot}$ is significantly smaller than the average mass of single CO-WDs ($\sim0.6\,M_{\rm \odot}$, Liebert et al. \cite{liebert05}) and very close to the tentative upper mass limit for WD companions to sdB stars seen in the sdB binary sample (Fig.~\ref{periodK}, see discussion in Kupfer et al. \cite{kupfer13}). Time-resolved photometry is needed to obtain a high-quality light curve of this system, perform a detailed analysis, and derive observational constraints on the mass-radius relation of the WD. Furthermore, sdB+WD binaries are important laboratories for studying relativistic effects such as Doppler boosting and microlensing (Geier et al. \cite{geier08}; Bloemen et al. \cite{bloemen11}; Geier et al. \cite{geier13}). 

OGLE\,BUL$-$SC16\,335 is the faintest HW\,Vir system known ($V\simeq16.5\,{\rm mag}$). It is located in the Galactic disc at a Galatic latitude of only $b=-3.5^{\rm \circ}$. Studying sdB binaries in different stellar populations is important for understanding their formation. Different metallicities and especially different ages are expected to  influence the properties of the progenitor binaries. Time-resolved spectroscopy and multicolour photometry are needed to perform a full analysis of OGLE\,BUL$-$SC16\,335, which was formed in the young disc population, and compare the results with the HW\,Vir systems found in the course of the MUCHFUSS project that originate from older populations like the thick disc or the halo. 

V\,1405\,Ori is one of only four short-period pulsators in a reflection effect binary (the others are NY\,Vir, Kilkenny et al. \cite{kilkenny98}; HE\,0230$-$4323, Kilkenny et al. \cite{kilkenny10}; 2M1938+4603, \O stensen et al. \cite{oestensen10a}; and FBS\,0117+396, \O stensen et al. \cite{oestensen13}). Such binaries are important as observational calibrators for current asteroseismic models of pulsating sdBs (e.g. van Grootel et al. \cite{vangrootel13}). Furthermore, the tidal influence of close companions might also influence the pulsational properties of the sdBs and should therefore be taken into account in the next generation of these models. 

\begin{acknowledgements}

A.T. was supported by the Deutsche Forschungsgemeinschaft (DFG) through grants HE1356/45-1. The research leading to these results has received funding from the European Research Council under the European Community's Seventh Framework Programme (FP7/2007--2013)/ERC grant agreement N$^{\underline{\mathrm o}}$\,227224 ({\sc prosperity}), from the Research Council of KU~Leuven grant agreement GOA/2008/04. We thank the referee Dave Kilkenny for his constructive report.

Based on observations at the Paranal Observatory of the European Southern Observatory for programmes number 165.H-0588(A) and 167.H-0407(A). Based on observations at the La Silla Observatory of the European Southern Observatory for programmes number 073.D-0495(A), 079.D-0288(A), 080.D-0685(A), 082.D-0649(A) and 086.D-0714(A). Based on observations collected at the Centro Astron\'omico Hispano Alem\'an (CAHA) at Calar Alto, operated jointly by the Max-Planck-Institut f\"ur Astronomie and the Instituto de Astrof\'isica de Andaluc\'ia (CSIC). Based on observations with the William Herschel Telescope operated by the Isaac Newton Group at the Observatorio del Roque de los Muchachos of the Instituto de Astrofisica de Canarias on the island of La Palma, Spain. Based on observations with the Southern Astrophysical Research (SOAR) telescope operated by the U.S. National Optical Astronomy Observatory (NOAO), the Ministério da Ciencia e Tecnologia of the Federal Republic of Brazil (MCT), the University of North Carolina at Chapel Hill (UNC), and Michigan State University (MSU). Based on observations at the McDonald observatory operated by the University of Texas in Austin. Based on observations made with the Mercator Telescope, operated
on the island of La Palma by the Flemish Community, at the Spanish Observatorio del Roque de los Muchachos of the Instituto de Astrofisica de Canarias.

\end{acknowledgements}

\newpage

\begin{appendix}

\vspace{10cm}
\section*{Appendix: Radial velocities}\label{app:RV}

\begin{center}
BPS\,CS\,22879$-$149\\
\begin{tabular}{lrl}
\hline\hline
\noalign{\smallskip}
mid$-$HJD & RV [${\rm km\,s^{-1}}$] & Instrument\\
$-2\,450\,000$ & & \\
\noalign{\smallskip}
\hline
\noalign{\smallskip}
4252.43920 &    2.6  $\pm$   9.0 & EMMI \\
4253.31834 &  -49.0  $\pm$   9.4 & \\
4253.40334 &   11.1  $\pm$  11.2 & \\
4254.25248 &  -47.9  $\pm$   9.0 & \\
4254.30548 &  -24.9  $\pm$   9.5 & \\
4254.36750 &   31.8  $\pm$   4.2 & \\
\noalign{\smallskip}
\hline
\noalign{\smallskip}
4756.50430  &   99.6  $\pm$  11.0  & EFOSC2 \\
5146.52218  &  -15.2  $\pm$   7.1  & \\
5147.51507  &   12.5  $\pm$   7.3  & \\
5147.52352  &     3.0 $\pm$   11.1 & \\
5147.52970  &    -1.3 $\pm$   6.3  & \\
5147.53589  &     6.5 $\pm$   8.3  & \\
5147.54742  &    33.6 $\pm$   7.0  & \\
5147.55499  &    37.9 $\pm$  11.6  & \\
5147.56256  &    42.8 $\pm$  11.0  & \\
5147.57460  &    40.4 $\pm$   8.1  & \\
5148.59060  &    79.5 $\pm$   8.5  & \\
5148.59678  &    74.4 $\pm$   8.5  & \\
5148.60296  &    82.1 $\pm$   7.7  & \\
5148.61428  &    77.9 $\pm$   8.0  & \\
5148.62717  &    86.3 $\pm$   8.5  & \\
\noalign{\smallskip}
\hline      
\noalign{\smallskip}
5412.82527  &  59.0  $\pm$  12.7   & Goodman \\ 
5412.82707  &  42.0  $\pm$  12.4   & \\ 
5412.82867  &  70.0  $\pm$  10.7   & \\ 
5412.89817  & -14.0  $\pm$  13.9    & \\
5412.89977  &   1.0  $\pm$  12.0   & \\ 
5412.90137  & -25.0  $\pm$  13.0    & \\
\noalign{\smallskip}
\hline
\end{tabular}
\end{center}

\begin{center}
HE\,1415$-$0309\\
\begin{tabular}{lrl}
\hline\hline
\noalign{\smallskip}
mid$-$HJD & RV [${\rm km\,s^{-1}}$] & Instrument\\
$-2\,450\,000$ & & \\
\noalign{\smallskip}
\hline
\noalign{\smallskip}
1740.63899 &  258.2   $\pm$   8.0 & UVES \\
1755.48571 &  203.1   $\pm$   8.0 & \\
\noalign{\smallskip}
\hline
\noalign{\smallskip}
4253.49757 &  149.0  $\pm$ 7.9   & EMMI \\
4253.63156 &  -86.9  $\pm$ 8.0   & \\
4253.70556 &  214.9  $\pm$ 29.6  & \\
4254.52251 &  185.0  $\pm$ 10.0  & \\
\noalign{\smallskip}
\hline
\noalign{\smallskip}
4476.85174  &  27.0  $\pm$ 8.6    & TWIN \\
4479.86504  & 253.0  $\pm$ 15.0   & \\
\noalign{\smallskip}
\hline      
\noalign{\smallskip}
4587.64673  &  -22.5  $\pm$ 7.8    & ISIS \\
\noalign{\smallskip}
\hline      
\noalign{\smallskip}
5240.78331  & 252.4  $\pm$  16.3   & Goodman \\ 
5240.79081  & 190.1  $\pm$   3.0   & \\ 
5240.79851  & 174.1  $\pm$   8.7   & \\ 
5240.80821  & 137.6  $\pm$   9.9    & \\
5240.81561  &  98.3  $\pm$  12.3   & \\ 
5240.82282  &  65.6  $\pm$  13.6    & \\
\noalign{\smallskip}
\hline
\end{tabular}
\end{center}

\newpage
\begin{center}
HS\,2043+0615\\
\begin{tabular}{lrl}
\hline\hline
\noalign{\smallskip}
mid$-$HJD & RV [${\rm km\,s^{-1}}$] & Instrument\\
$-2\,450\,000$ & & \\
\noalign{\smallskip}
\hline
\noalign{\smallskip}
2387.90703  &  27.2    $\pm$ 5.0  & UVES \\
2521.66074  &  -108.9  $\pm$ 5.0  & \\
\noalign{\smallskip}
\hline
\noalign{\smallskip}
4586.73051 &  -19.2  $\pm$ 11.0  & ISIS \\
4586.73799 &  -31.0  $\pm$ 10.9  & \\
\noalign{\smallskip}
\hline
\noalign{\smallskip}
4692.49332  &   39.2  $\pm$ 10.0 & TWIN \\
4693.48658  &  -51.9  $\pm$  7.0 & \\
4693.51664  &  -83.2  $\pm$  6.0 & \\
4694.47290  & -119.5  $\pm$  4.0 & \\
4696.57722  &  -97.3  $\pm$  6.0 & \\
4696.62917  &  -78.9  $\pm$  6.0 & \\
\noalign{\smallskip}
\hline
\noalign{\smallskip}
4756.53557  &   -71.6 $\pm$ 21.0 & EFOSC2 \\
4757.52335  &  -121.4 $\pm$ 13.0 & \\
4757.63634  &    27.6 $\pm$ 16.0 & \\
4758.66733  &  -116.3 $\pm$ 17.0 & \\
4758.67470  &  -133.9 $\pm$ 15.0 & \\
\noalign{\smallskip}
\hline
\end{tabular}
\end{center}

\vfill
\begin{center}
HS\,2359+1942\\
\begin{tabular}{lrl}
\hline\hline
\noalign{\smallskip}
mid$-$HJD & RV [${\rm km\,s^{-1}}$] & Instrument\\
$-2\,450\,000$ & & \\
\noalign{\smallskip}
\hline
\noalign{\smallskip}
2610.54512 &    4.4  $\pm$   2.0 & UVES \\
\noalign{\smallskip}
\hline
\noalign{\smallskip}
4252.91279 &    57.6  $\pm$   20.0 & EMMI \\
\noalign{\smallskip}
\hline
\noalign{\smallskip}
5068.69977  & -87.7  $\pm$ 4.5  & ISIS \\
5068.70011  & -82.6  $\pm$ 4.5  & \\
5068.70725  & -78.3  $\pm$ 3.8  & \\
5068.71712  & -65.2  $\pm$ 4.6  & \\
5068.72433  & -62.6  $\pm$ 5.0  & \\
5068.73153  & -60.0  $\pm$ 4.4  & \\
5068.74844  & -53.9  $\pm$ 4.6  & \\
5068.75565  & -39.0  $\pm$ 4.0  & \\
5069.66466  & -80.2  $\pm$ 4.8  & \\
5069.67188  & -79.5  $\pm$ 6.0  & \\
5069.67909  & -63.9  $\pm$ 4.4  & \\
5070.74702  &   2.7  $\pm$ 5.3  & \\
5070.75423  &   3.8  $\pm$ 5.2  & \\
5070.76144  &   1.7  $\pm$ 3.3  & \\
\noalign{\smallskip}
\hline      
\noalign{\smallskip}
4755.63418  &   -1.7  $\pm$ 29.2  & EFOSC2 \\
4755.64807  &  -27.0  $\pm$ 13.0  & \\
4756.73373  & -139.1  $\pm$ 17.5  & \\
4758.70798  & -164.3  $\pm$ 16.9  & \\
4758.71068  & -156.7  $\pm$ 16.4  & \\
\noalign{\smallskip}
\hline      
\noalign{\smallskip}
5412.83585  &  -88.1  $\pm$  12.9   & Goodman \\ 
5412.85165  &  -90.6  $\pm$  7.5   & \\ 
\noalign{\smallskip}
\hline      
\noalign{\smallskip}
6277.39152  & -71.1  $\pm$   6.2    & ISIS \\ 
6277.39885  & -57.8  $\pm$   2.1     & \\
6277.40252  & -60.9  $\pm$   5.0    & \\ 
6279.48917  &  11.8  $\pm$   5.1    & \\
6279.49283  & 14.4   $\pm$   6.2   & \\ 
6279.49650  & 14.7   $\pm$   7.8    & \\
6279.50016  & 18.2   $\pm$   2.1   & \\ 
6280.40209  &  20.5  $\pm$   8.0     & \\
6280.40576  &  10.4  $\pm$   5.3    & \\ 
6280.40942  &   6.1  $\pm$   5.0     & \\
6280.41309  &   7.8  $\pm$   5.0    & \\ 
6280.43805  &   6.0  $\pm$  26.7     & \\
6280.44171  &   8.6  $\pm$   5.5     & \\
6280.44538  &   2.8  $\pm$   5.0    & \\ 
6280.47358  & -17.4  $\pm$   6.5     & \\
\noalign{\smallskip}
\hline
\end{tabular}
\end{center}
\vfill

\begin{center}
LB\,1516\\
\begin{tabular}{lrl}
\hline\hline
\noalign{\smallskip}
mid$-$HJD & RV [${\rm km\,s^{-1}}$] & Instrument\\
$-2\,450\,000$ & & \\
\noalign{\smallskip}
\hline
\noalign{\smallskip}
1795.62809   & -25.3   $\pm$ 2.0 & FEROS \\
1795.65378   & -22.7   $\pm$ 2.0 & \\
2495.87673   &  35.0   $\pm$ 2.0 & \\
2497.86693   & -20.4   $\pm$ 2.0 & \\
3250.67647   &  63.3   $\pm$ 2.0 & \\
3251.57092   &  49.2   $\pm$ 2.0 & \\
3253.55317   &  -5.8   $\pm$ 2.0 & \\
3253.67530   &  -7.5   $\pm$ 2.0 & \\ 
\noalign{\smallskip}
\hline      
\noalign{\smallskip}
4755.83240   &  -30.2  $\pm$  8.8  & EFOSC2 \\ 
4756.74952   &  -44.5  $\pm$  8.0  & \\
4758.77238   &  -12.7  $\pm$ 10.0  & \\
4758.77510   &  -11.9  $\pm$  9.0  & \\
\noalign{\smallskip}
\hline      
\noalign{\smallskip}
5499.57381   &  62.0  $\pm$  2.0  & FEROS \\
5499.59649   &  44.0  $\pm$  2.0  & \\
\noalign{\smallskip}
\hline
\end{tabular}
\end{center}

\begin{center}
OGLE\,BUL$-$SC16\,335\\
\begin{tabular}{lrl}
\hline\hline
\noalign{\smallskip}
mid$-$HJD & RV [${\rm km\,s^{-1}}$] & Instrument\\
$-2\,450\,000$ & & \\
\noalign{\smallskip}
\hline
\noalign{\smallskip}
4757.58903  &  -23.9  $\pm$ 13.0 & EFOSC2 \\
4758.49724  &   55.5  $\pm$ 13.0 & \\
4758.50460  &  103.8  $\pm$ 16.0 & \\
4758.59169  &  -57.8  $\pm$ 15.0 & \\
4758.60219  &    2.2  $\pm$ 17.0 & \\
4758.61866  &   36.1  $\pm$ 28.0 & \\
\noalign{\smallskip}
\hline
\end{tabular}
\end{center}

\begin{center}
PG\,0941+280\\
\begin{tabular}{lrl}
\hline\hline
\noalign{\smallskip}
mid$-$HJD & RV [${\rm km\,s^{-1}}$] & Instrument\\
$-2\,450\,000$ & & \\
\noalign{\smallskip}
\hline
\noalign{\smallskip}
3715.01112  & 210.4  $\pm$ 1.9 & McDonald \\
3716.99883  & -60.2  $\pm$ 2.3 & \\
3717.02115  & -61.1  $\pm$ 4.1 & \\
3766.89779  & 106.3  $\pm$ 3.7 & \\
3768.86376  & 211.5  $\pm$ 3.2 & \\
\noalign{\smallskip}
\hline
\noalign{\smallskip}
4476.76052  &  -3.3 $\pm$ 9.2  & EMMI \\
4476.84388  & 184.8 $\pm$ 13.5 & \\
4477.78158  & 190.7 $\pm$ 6.5  & \\
4478.75103  & 190.3 $\pm$ 6.8  & \\
4479.71960  & 174.8 $\pm$ 13.8 & \\
4479.82016  & -45.1 $\pm$ 9.2  & \\
\noalign{\smallskip}
\hline
\noalign{\smallskip}
4979.36163  & 133.3 $\pm$ 12.3 & TWIN \\
4979.41243  &  -1.0 $\pm$ 10.9 & \\
4980.36049  & -41.0 $\pm$ 10.0 & \\
4980.41073  & -65.8 $\pm$ 11.2 & \\
4980.42660  & -55.0 $\pm$ 11.7 & \\
4981.40285  &  29.2 $\pm$ 10.7 & \\
\noalign{\smallskip}
\hline
\end{tabular}
\end{center}

\vfill
\begin{center}
V\,1405\,Ori\\
\begin{tabular}{lrl}
\hline\hline
\noalign{\smallskip}
mid$-$HJD & RV [${\rm km\,s^{-1}}$] & Instrument\\
$-2\,450\,000$ & & \\
\noalign{\smallskip}
\hline
\noalign{\smallskip}
4476.68258  &   17.3  $\pm$ 10.3 & EMMI \\
4477.59921  &  -68.1  $\pm$ 18.4 & \\
4477.64791  & -129.6  $\pm$ 13.4 & \\
4477.69907  & -108.1  $\pm$ 31.6 & \\
4478.58382  &    9.3  $\pm$ 17.0 & \\
4478.64047  &   63.7  $\pm$  8.6 & \\
4478.70892  &   26.3  $\pm$ 14.5 & \\
4479.61084  & -117.8  $\pm$  7.9 & \\
4479.75802  &  -40.1  $\pm$  9.3 & \\
\noalign{\smallskip}
\hline
\noalign{\smallskip}
4692.65549 & -58.1 $\pm$  9.0 & TWIN \\
4692.66317 & -37.8 $\pm$ 10.0 & \\
4692.67134 & -15.8 $\pm$ 10.0 & \\
4692.67896 &  94.1 $\pm$ 12.0 & \\
4693.69611 & -49.5 $\pm$ 11.0 & \\
4696.66314 &  -4.6 $\pm$  3.0 & \\
\noalign{\smallskip}
\hline
\end{tabular}
\end{center}

\end{appendix}


\begin{thebibliography}{}

\bibitem[2009]{abazajian09}
Abazajian, K. N., Adelman-McCarthy, J. K., Ag\"ueros, M. A., et al. 2009, ApJS, 182, 543

\bibitem[1998]{baraffe98}
Baraffe, I., Chabrier, G, Allard, F., \& Haunschildt, P. H. 1998, A\&A, 337, 403

\bibitem[1992]{beers92}
Beers, T. C., Preston, G. W., Shectman, S. A., Doinidis, S. P., \& Griffin, K. E. 1992, AJ, 103, 267

\bibitem[2011]{bloemen11}
Bloemen, S., Marsh, T. R., \O stensen, R. H., et al. 2011, MNRAS, 410, 1787

\bibitem[2011]{classen11}
Classen, L., Geier, S., Heber, U., \& O'Toole, S. J. 2011, AIP Conf. Ser., 1331, 297

\bibitem[2010]{copperwheat11} 
Copperwheat, C., Morales-Rueda, L., Marsh, T. R., et al. 2011, MNRAS, 415, 1381

\bibitem[2005]{edelmann05}
Edelmann, H., Heber, U., Altmann, M., Karl, C., \& Lisker, T. 2005a, A\&A 442, 1023

\bibitem[2012]{fontaine12}
Fontaine, G., Brassard, P., Charpinet, S., et al. 2012, A\&A, 539, 12

\bibitem[2007]{geier07}
Geier, S., Nesslinger, S., Heber, U., et al. 2007, A\&A, 464, 299

\bibitem[2008]{geier08}
Geier, S., Nesslinger, S., Heber, U., et al. 2008, A\&A, 477, L13

\bibitem[2010a]{geier10a}
Geier, S., Heber, U., Kupfer, T., \& Napiwotzki, R. 2010a, A\&A, 515, 37

\bibitem[2010b]{geier10b}
Geier, S., Heber, U., Podsiadlowski, Ph., et al. 2010b, A\&A, 519, 25

\bibitem[2011a]{geier11a} 
Geier, S., Hirsch, H., Tillich, A., et al. 2011a, A\&A, 530, 28

\bibitem[2011b]{geier11b} 
Geier, S., Maxted, P. F. L., Napiwotzki, R., et al. 2011b, A\&A, 526, 39

\bibitem[2011c]{geier11c} 
Geier, S., Schaffenroth, V., Drechsel, H., et al. 2011c, ApJ, 731, L22

\bibitem[2012]{geier12}
Geier, S., Schaffenroth, V., Hirsch, H., et al. 2012, ASP Conf. Ser., 452, 129

\bibitem[2013]{geier13}
Geier, S., Marsh, T, R., Wang, B., et al. 2013, A\&A, 554, 54

\bibitem[2004]{green04}
Green, E. M., For, B., Hyde, E. A., et al. 2004, Ap\&SS, 291, 267

\bibitem[2002]{han02}
Han Z., Podsiadlowski P., Maxted P. F. L., Marsh T. R., \& Ivanova N. 2002, 
MNRAS, 336, 449

\bibitem[2003]{han03}
Han, Z., Podsiadlowski, P., Maxted, P. F. L., \& Marsh, T. R. 2003, MNRAS, 341, 669

\bibitem[1986]{heber86}
Heber, U. 1986, A\&A, 155, 33

\bibitem[2009]{heber09} 
Heber, U. 2009, ARA\&A, 47, 211 

\bibitem[1998]{kilkenny98}
Kilkenny, D., O’Donoghue, D., Koen, C., Lynas-Gray, A. E., \& van Wyk, F. 1998, MNRAS, 296, 329

\bibitem[2010]{kilkenny10}
Kilkenny, D., Koen, C., \& Worters, H. 2010, MNRAS, 404, 376

\bibitem[1999]{koen99}
Koen, C., O'Donoghue, D., Kilkenny, D., Stobie, R. S., \& Saffer, R. A. 1999, MNRAS, 306, 213

\bibitem[2010]{koen10}
Koen, C., Kilkenny, D., Pretorius, M. L., \& Frew, D. J. 2010, MNRAS, 401, 1850

\bibitem[2013]{kupfer13}
Kupfer, T., Geier, S., Barlow, B. N., et al. 2013, A\&A, submitted

\bibitem[2005]{liebert05}
Liebert, J., Bergeron, P., \& Holberg, J. B. 2005, ApJS, 156, 47

\bibitem[2005]{lisker05}
Lisker, T., Heber, U., Napiwotzki, R., Christlieb, N., Han, Z., et al. 2005, A\&A, 430, 223

\bibitem[2000]{maxted00}
Maxted, P. F. L., Marsh, T. R., \& North, R. C. 2000, MNRAS, 317, L41

\bibitem[2001]{maxted01}
Maxted, P. F. L., Heber, U., Marsh, T. R., \& North, R. C. 2001, MNRAS, 326, 139 

\bibitem[2003]{morales03}
Morales-Rueda, L., Maxted, P. F. L., Marsh, T. R., North, R. C., \& Heber, U. 2003, MNRAS, 338, 752

\bibitem[2003]{napiwotzki03}
Napiwotzki, R., Christlieb, N., Drechsel, H., et al. 2003, ESO Msngr, 112, 25

\bibitem[2004a]{napiwotzki04a}
Napiwotzki, R., Karl, C., Lisker, T., et al. 2004a, Ap\&SS, 291, 321

\bibitem[2004b]{napiwotzki04b}
Napiwotzki, R., Yungelson, L., Nelemans, G. et al. 2004b, ASP Conf. Ser., 318, 402

\bibitem[2010]{nelemans10}
Nelemans, G. 2010, Ap\&SS, 329, 25

\bibitem[2010]{oestensen10}
\O stensen, R. H. 2010, AN, 331, 1026

\bibitem[2010a]{oestensen10a}
\O stensen, R. H., Green, E. M., Bloemen, S., et al. 2010a, MNRAS, 408, L51

\bibitem[2010b]{oestensen10b}
\O stensen, R. H., Oreiro, R., Solheim, J.-E., et al. 2010b, A\&A, 513, 6

\bibitem[2013]{oestensen13}
\O stensen, R. H., Geier, S., Schaffenroth, V., et al. 2013, A\&A, 559, 35

\bibitem[1999]{orosz99}
Orosz, J. A., \& Wade, R. A. 1999, MNRAS, 310, 773

\bibitem[2006]{otoole06}
O'Toole, S. J., \& Heber, U. 2006, A\&A, 452, 579

\bibitem[2003]{pfahl03}
Pfahl, E., Rappaport, S., \& Podsiadlowski, Ph. 2003, ApJ, 597, 1036

\bibitem[2002]{podsi02}
Podsiadlowski, Ph., Rappaport, S., \& Pfahl, E. D. 2002, ApJ, 565, 1107

\bibitem[2007]{polubek07}
Polubek, G., Pigulski, A., Baran, A., \& Udalski, A. 2007, ASP Conf. Ser., 372, 487

\bibitem[2013]{raskin13}
Raskin, G., Bloemen, S., Morren, J., et al. 2013, A\&A, 559, 26

\bibitem[2010]{reed10}
Reed, M. D., Terndrup, D. M., \O stensen, R. H., et al. 2010, Ap\&SS, 329, 83

\bibitem[1994]{saffer94}
Saffer, R. A., Bergeron, P., Koester, D., \& Liebert, J. 1994, ApJ, 432, 351

\bibitem[2013a]{schaffenroth13a}
Schaffenroth, V., Geier, S., Heber, U., et al. 2013a, A\&A, submitted

\bibitem[2013b]{schaffenroth13b}
Schaffenroth, V., Geier, S., Ziegerer, E., et al. 2013b, A\&A, submitted

\bibitem[2012]{telting12}
Telting, J. H., \O stensen, R. H., Baran, A. S., et al. 2012, A\&A, 544, 1

\bibitem[2011]{tillich11}
Tillich, A., Heber, U., Geier, S., et al. 2011, A\&A, 527, 137

\bibitem[1981]{tutukov81}
Tutukov, A. V., \& Yungelson, L. R. 1981, Nauchnye Informatsii, 49, 3

\bibitem[2013]{vangrootel13}
van Grootel. V., Charpinet, S., Brassard, P., Fontaine, G., \& Green, E. M. 2013, A\&A, 553, 97

\bibitem[2012]{vennes12} 
Vennes, S., Kawka, A., O'Toole, S. J., N\'emeth, P, \& Burton, D. 2012, ApJ, 759, L25

\bibitem[2005]{yungelson05}
Yungelson, L. R., \& Tutukov, A. V. 2005, ARep, 49, 871

\bibitem[2013]{wang13}
Wang, B., Justham, S., \& Han, Z. 2013, A\&A, 559, 94

\bibitem[1984]{webbink84} 
Webbink, R.~F.\ 1984, ApJ, 277, 355 

\end{thebibliography}
\end{document}